\documentclass[fleqn,usenatbib]{mnras}

\usepackage{newtxtext,newtxmath}

\usepackage[T1]{fontenc}
\usepackage{ae,aecompl}

\usepackage{graphicx}	
\usepackage{amsmath}	
\usepackage{ulem}       
\usepackage[dvipsnames]{xcolor} 

\newcommand{\arnps}{Ann.\,Rev.\,Nuc.\,Par.\,Sci.}
\newcommand{\cac}{ComAC}
\newcommand\jpgnp{J.\,Phys.\,G\,\,Nucl.\,Phys.}
\newcommand{\cmpers}{\,\mathrm{cm}\,\mathrm{s}^{-1}\,}
\newcommand{\kmpers}{\,\mathrm{km}\,\mathrm{s}^{-1}\,}
\newcommand{\ergpers}{\,\mathrm{erg}\,\mathrm{s}^{-1}\,}
\newcommand{\radpers}{\,\mathrm{rad}\,\mathrm{s}^{-1}\,}
\newcommand{\gperccm}{\,\mathrm{g}\,\mathrm{cm}^{-3}\,}
\newcommand{\kbpbaryon}{\,k_{B}\,\mathrm{baryon}^{-1}\,}
\newcommand{\mathkm}{\,\mathrm{km}\,}
\newcommand{\lnu}{L_{\nu}}

\newcommand{\ellc}{\ell_{c}}
\newcommand{\lcon}{L_{\mathrm{con}}}

\newcommand{\lwavel}{L_{\mathrm{waves},\,\ell}}
\newcommand{\ltransl}{L_{\mathrm{trans},\,\ell}}
\newcommand{\lheatl}{L_{\mathrm{heat},\,\ell}}
\newcommand{\lwave}{L_{\mathrm{waves}}}
\newcommand{\ltrans}{L_{\mathrm{trans}}}
\newcommand{\lheat}{L_{\mathrm{heat}}}
\newcommand{\msun}{M_{\sun}}

\newcommand{\vcon}{v_{\mathrm{con}}}
\newcommand{\wcon}{\omega_{\mathrm{con}}}
\newcommand{\rcon}{r_{\mathrm{con}}}
\newcommand{\drcon}{\Delta\,\rcon}
\newcommand{\Mcon}{\mathcal{M}_{\mathrm{con}}}
\newcommand{\rshock}{r_{\mathrm{shock}}}

\title[Wave heating in supernovae]{Wave heating from proto-neutron star convection and the core-collapse supernova explosion mechanism}

\author[S. E. Gossan et al.]{
Sarah E. Gossan,$^{1,2,3}$\thanks{E-mail: gossan@cita.utoronto.ca}
Jim Fuller,$^{2}$
and Luke F. Roberts$^{4}$
\\
$^{1}$Canadian Institute for Theoretical Astrophysics, University of Toronto, Toronto, ON M5S 3H8, Canada\\
$^{2}$TAPIR, Mailcode 350-17, California Institute of Technology, Pasadena, CA 91125, USA\\
$^{3}$LIGO Lab, Mailcode 100-36, California Institute of Technology, 1200 E California Blvd, Pasadena, CA 91125, USA\\
$^{4}$NSCL and Department of Physics and Astronomy, Michigan State University, East Lansing, MI 48824, USA
}

\date{Accepted 2019 November 18. Received 2019 November 15; in original form 2019 October 16}

\pubyear{2019}

\begin{document}
\label{firstpage}
\pagerange{\pageref{firstpage}--\pageref{lastpage}}
\maketitle

\begin{abstract}
Our understanding of the core-collapse supernova explosion mechanism is 
incomplete. While the favoured scenario is delayed revival of the stalled shock 
by neutrino heating, it is difficult to reliably compute explosion outcomes and 
energies, which depend sensitively on the complex radiation hydrodynamics of 
the post-shock region. The dynamics of the (non-)explosion depend sensitively 
on how energy is transported from inside and near the proto-neutron star (PNS) 
to material just behind the supernova shock.  Although most of the PNS energy is 
lost in the form of neutrinos, hydrodynamic and hydromagnetic waves can also 
carry energy from the PNS to the shock. We show that gravity waves excited by 
core PNS convection can couple with outgoing acoustic waves that present an 
appreciable source of energy and pressure in the post-shock region. Using 
one-dimensional simulations, we estimate the gravity wave energy flux excited 
by PNS convection and the fraction of this energy transmitted upward to the 
post-shock region as acoustic waves. We find wave energy fluxes near 
$10^{51}\ergpers$ are likely to persist for $\sim \! 1\,\mathrm{s}$ 
post-bounce. The wave pressure on the shock may exceed 10 per cent of the thermal 
pressure, potentially contributing to shock revival and, subsequently, a
successful and energetic explosion. We also discuss how future simulations can 
better capture the effects of waves, and more accurately quantify wave heating 
rates.             
\end{abstract}




\section{Motivation}
\label{sec:motivation}
The core-collapse supernova (CCSN) explosion mechanism is not well understood. 
Several mechanisms have been proposed 
(see, e.g.,~\citealt{janka:2012,janka:etal:2016,bmueller:2016,burrows:etal:2018}  
for broad reviews), but delayed shock revival through neutrino heating is       
favoured for progenitor cores with pre-collapse rotational periods greater than 
a few seconds~\citep{bethe:wilson:1985}. The ability of the so-called delayed 
neutrino-heating mechanism to robustly drive energetic supernova explosions in a wide 
range of progenitor stars has yet to be fully established, although a subset 
of three-dimensional simulations are beginning to predict successful 
explosions for various progenitor stars~\citep{janka:etal:2016, ott:etal:2018, burrows:19}. What has become 
clear is that the development and sustained presence of multi-dimensional 
hydrodynamic instabilities in the post-shock accretion flow, such as turbulent 
neutrino-driven convection and the standing accretion-shock instability         
(SASI; see, e.g.,~\citealt{foglizzo:etal:2007} and references therein) is crucial 
for successful explosions. Even when explosions are produced, it is often not   
clear what the total explosion energy should be, nor how it should be dependent 
on progenitor properties or the explosion dynamics.

The success or failure of the delayed neutrino-heating mechanism can be
sensitive to small changes in the physics included in CCSN simulations, 
particularly to physics that effect the properties of the proto-neutron star 
(PNS;~\citealt{melson:etal:2015:b, bollig:etal:2017, burrows:etal:2018, schneider:etal:2019}). 
Neutrino emission energising the shock in the pre-explosion phase is powered 
predominantly by accretion onto the PNS~\citep{bmueller:janka:2014}, but this 
same accretion provides ram pressure which prevents shock runaway. This balance 
can be tilted in favour of explosion if the PNS is able to more 
efficiently couple the accretion power back to the post-shock region, for 
instance by increasing the average energy of the emitted 
neutrinos~\citep[e.g.][]{janka:etal:2016}.  Although accretion power provides 
the majority of the flux in electron neutrinos and anti-electron neutrinos 
prior to explosion, the potential for 
gravitational contraction of the inner regions of the PNS provides a huge 
reservoir of energy, totalling well over $10^{53}\,\mathrm{erg}$. Due to the 
relatively long diffusion timescale of neutrinos through the PNS, energy 
released through core contraction contributes a fairly small fraction of 
the total luminosity during the pre-explosion phase. Nevertheless, only 
$\sim \! 5$ per cent of the total neutrino luminosity is deposited in the
gain region behind the stalled shock. Should another mechanism operate, one capable 
of transporting energy from the gravitational reservoir in the PNS core and 
depositing it in the post-shock region more efficiently than neutrinos, the 
energy released via gravitational contraction of the core could help to 
facilitate the delayed neutrino-heating mechanism.

Unsurprisingly, the long-term multineutrino energy group radiation 
hydrodynamic simulations required to explore the explosion mechanism continue 
to present a computationally daunting task. State-of-the-art simulations using 
three-flavour, multigroup neutrino radiation transport in multiple dimensions 
(see, e.g.,~\citealt{lentz:etal:2015,melson:etal:2015:b,janka:etal:2016,roberts:etal:2016,vartanyan:etal:2019,burrows:etal:2019:a})
have seen some successful explosions, although less energetic than expected. 
While failed explosions are still seen with these more involved simulations, 
several studies have shown that the evolutionary track for many models exists 
near the boundary between successful explosion and failed supernova. It has 
been shown that modest changes to input physics and numerical techniques such 
as using more realistic input data for the progenitor models, can 
reduce the critical neutrino luminosity required for a successful explosion to 
develop~\citep{couch:etal:2015,abdikamalov:etal:2016,abdikamalov:etal:2018}.

In the interest of reducing computational costs, a long-favoured tactic has     
been to employ a more coarsely spaced (i.e. lower resolution) radial grid. It 
has been shown, however, that failure to resolve turbulence across              
the inertial range of spatial scales reduces the turbulent pressure beneath the 
stalled shock, unintentionally further inhibiting 
explosions~\citep{abdikamalov:etal:2015,radice:etal:2015,couch:ott:2015}. 
Additionally, this low resolution may impact the development of convection 
within the PNS. Another strategy used in some simulations is evolving part of 
the PNS in spherical 
symmetry~\citep[e.g.][]{hanke:etal:2012,bmueller:2015,bmueller:etal:2017},
which increases the Courant limit on the time-step. The elimination of 
non-radial hydrodynamics in the PNS, however, may suppress the development of 
convection there. PNS convection may efficiently transport energy 
out from the central regions of the PNS to near the neutrino decoupling radius, 
potentially increasing the emitted neutrino 
luminosities~\citep[e.g.][]{dessart:etal:2006:b} and exciting outgoing internal 
gravity waves above the PNS.

Generally speaking, the impact of the approximations highlighted above on the   
``explodability'' of realistic progenitor stars is only beginning to be 
understood. As a consequence, it is important to investigate the significance 
of additional physical processes (or ``mechanisms'') contributing to the 
dynamics of shock revival, even if they are sub-dominant.

In this paper, we seek to explore the effect of heating from gravito-acoustic 
waves excited by PNS convection on shock revival and explosion energy in the 
context of CCSNe. In the following, we outline the basic concept of our 
proposed idea in Section~\ref{sec:basicidea}, detail the one-dimensional 
simulations employed for this study in Section~\ref{sec:sims}, and describe our 
calculations of the wave energy fluxes in Section~\ref{sec:wavegen}. In 
Section~\ref{sec:damping-and-non-lin}, we examine wave damping processes and 
the impact of non-linear wave dynamics, before discussing the implications 
(and limitations) of our work in light of these effects in 
Section~\ref{sec:discussion}, and concluding with Section~\ref{sec:conclusion}.

\section{Basic idea}
\label{sec:basicidea}
The evolution of the PNS and supernova shock wave during the first few hundred  
milliseconds after core bounce is, for the most part, agreed upon by            
simulations~\citep{oconnor:etal:2018}. The shock's energy 
is drained through a combination of neutrino losses and photodissociation of 
infalling heavy nuclei, and the shock wave stalls (at radius $\rshock$) between 
$150$ and $250\mathkm$. The subsonic material interior to   
(beneath) the shock is roughly in hydrostatic equilibrium. Within 
$\sim \! 100\,\mathrm{ms}$ of core bounce, regions of net neutrino heating 
(in the post-shock ``gain region'') and net neutrino cooling (above the PNS 
neutrinosphere, roughly located around density 
$\rho\sim10^{11}\gperccm$) develop. The negative entropy
gradient that emerges in the gain region drives vigorous convection there which 
rapidly becomes turbulent. The gain radius, marking the inner bound of the net 
heating region beneath the stalled shock, is typically between $50$ 
and $100\mathkm$. Beneath the gain region exists a radiative layer that 
is stably stratified through net neutrino cooling that creates a positive 
entropy gradient. This layer extends below the neutrinosphere, interior to 
which neutrinos are strongly coupled to (i.e. trapped by) the dense nuclear 
matter comprising the inner PNS core. Meanwhile, gradients in composition and 
entropy develop between the hot, lepton-rich core and the neutrino-cooled, 
deleptonized neutrinospheres, driving convection in the PNS mantle. Many 
simulations confirm the development a convective layer in the PNS mantle within 
$150$ to $200\,\mathrm{ms}$ after core bounce, at radii between 
roughly $10$ and $20\mathkm$~\citep[e.g.][]{dessart:etal:2006:b}.

\begin{figure}
\centering
\includegraphics[width=0.97\columnwidth]{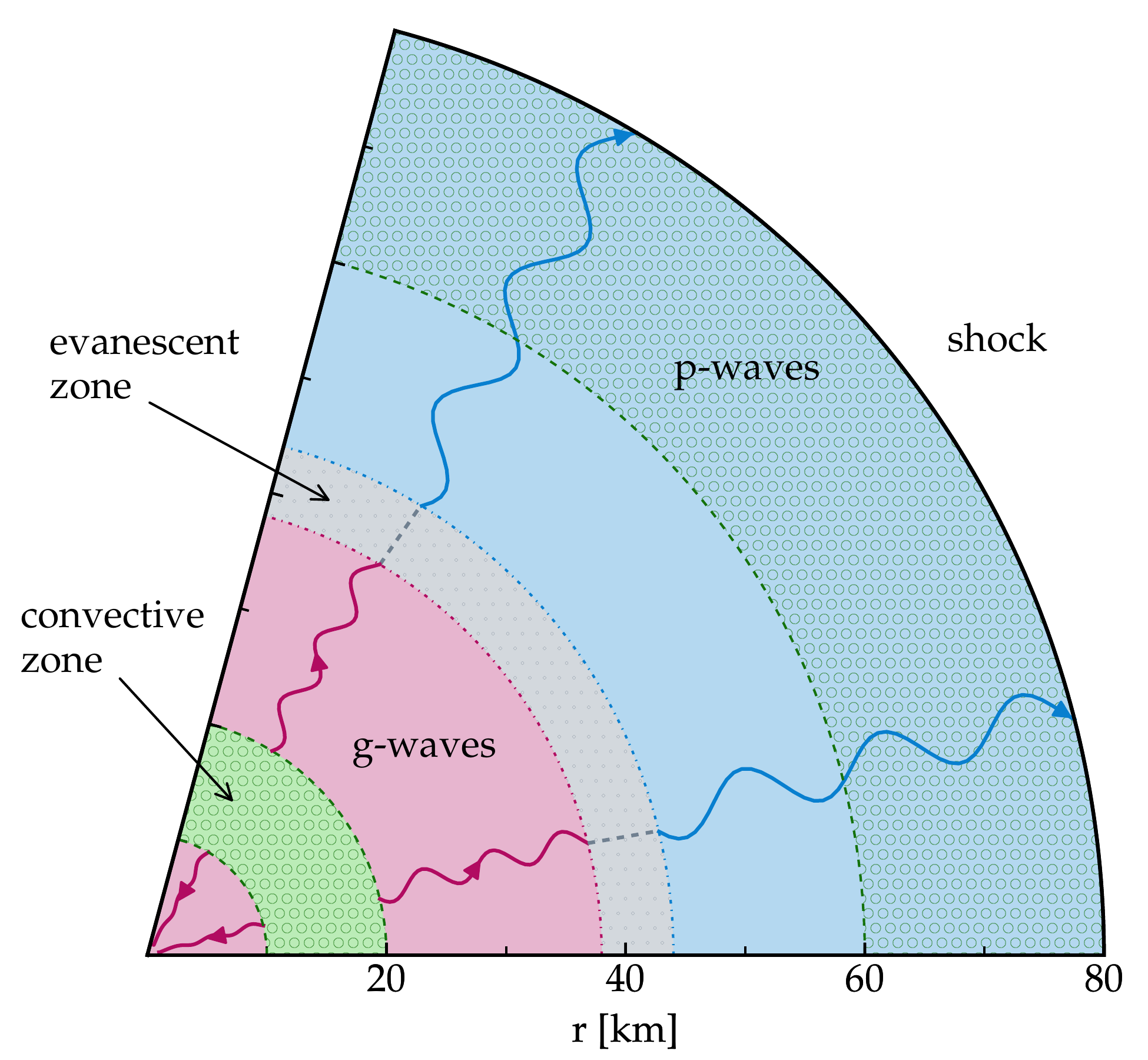}
\caption{
Gravity waves are excited by turbulent convection (green hatched regions) in the PNS and propagate outward (pink regions). At evanescent zones (grey region), the waves either reflect back towards the PNS, or they tunnel through the evanescent region to generate outgoing acoustic waves. These waves propagate outwards until they are damped or they encounter the shock, beyond which they cannot propagate. Both wave pressure on the shock and dissipation of wave energy in the gain region may aid shock revival.}
\label{fig:diagram}
\end{figure}


We are concerned with the effects of energy transport via hydrodynamic waves    
from the PNS mantle out to the post-shock region. As discussed              
by~\citet{goldreich:kumar:1990}, internal gravity waves are generated by    
turbulent convection and emitted at convective-radiative interfaces,            
transporting energy and angular momentum away from the convective zone, and 
depositing it where the waves damp away. Energy and angular momentum transport  
through gravity waves generated via this mechanism can be significant, and has  
been shown to have important ramifications inside low-mass, Sun-like stars      
(e.g.~\citealt{kumar:etal:1999,talon:etal:2002,fuller:etal:2014}),          
intermediate-mass stars (e.g.~\citealt{rogers:etal:2012,rogers:etal:2013}), 
and massive stars (see, 
e.g.,~\citealt{meakin:arnett:2007,quataert:shiode:2012,fuller:etal:2015:a,fuller:2017}).

In the context of the PNS and the evolving supernova shock, gravity waves are   
expected to be excited in two regions; the optically thick and convective PNS 
mantle, and secondly in the gain region below the shock due to
neutrino-driven convection. The detailed study of~\citet{dessart:etal:2006:b} 
(see also~\citealt{yoshida:etal:2007}) examines the properties of the inner PNS 
convection zone and the gravity waves in the overlying radiative                
(neutrino-cooled) layer. They find gravity waves with angular index $\ell=1-3$  
and frequencies of $\omega\sim10^{3}\radpers$ are        
prevalent, though they attribute their excitation primarily to the (more        
vigorous) convection in the outer convective zone. We note, however, that       
gravity waves are also excited by the convection zone in the PNS mantle.        
Gravity waves generated here will couple to acoustic waves in the outer PNS,    
where they then propagate out towards the stalled shock. Energy and angular     
momentum is transported outwards by these waves from the convective zone in the 
PNS mantle (which is driven, ultimately, by the release of gravitational binding
energy as the nascent remnant cools and contracts over the time-scale of a few   
seconds), and deposited in regions of heavy damping. Should this damping occur  
predominantly in the gain region, the increased energy deposition may augment   
the thermal pressure behind the shock, and aid in its revival. Alternatively,   
waves that propagate all the way to the shock will be reflected back inwards, 
contributing an additional source of pressure upon the shock to drive it 
outwards. We illustrate this scenario in Fig.~\ref{fig:diagram} with a 
stylised cartoon diagram.

We note that this process is related to the ``acoustic mechanism'' proposed     
by~\citet{burrows:etal:2006:a}, but with key differences. While the former is   
powered by energy from accretion of infalling material onto the outer PNS, the  
acoustic energy we consider comes from the liberation of gravitational binding  
energy as the PNS core deleptonizes and contracts. This reservoir, containing 
well over $10^{53}\,\mathrm{erg}$ of binding energy at capacity, remains 
tappable long after accretion onto the PNS has ceased. Consequently, gravity 
waves excited in this way have the potential to aid shock revival and help to 
drive the explosion for several seconds after core bounce. The transport of a 
small fraction of the PNS binding energy out to the gain region via gravity 
waves may be considered an aspect of the so-called CCSN ``central engine". It 
follows that effects from these waves may be missed in simulations that (for 
the purposes of reducing computational cost) do not follow PNS evolution either 
(i) long enough to witness the development of a convective layer in the PNS 
mantle, or (ii) with sufficient degrees of freedom in long-running simulations, 
suppressing the development of non-radial hydrodynamic instabilities and, in 
turn, their effect. Even in simulations that may capture these waves, their 
impact on the explosion mechanism has not been explicitly noted and quantified.

First and foremost, the goal of the study presented here is to determine the 
extent to which gravito-acoustic waves generated by PNS convection could 
contribute to shock revival. This could increase the fraction of 
explosions seen by simulations evolving CCSNe, as well as the characteristic 
explosion energies. We employ spherically symmetric simulations to estimate 
the spectral behaviour and energy flux of waves excited 
by PNS convection, then quantitatively estimate heating rates in the gain 
region from wave energy transport for the first $660\,\mathrm{ms}$ after core 
bounce. We discuss the assumptions made to calculate these estimates, and 
explore how wave damping and non-linear effects could impact our results. 
Lastly, we consider how wave dynamics can be captured and heating rates 
quantified in future simulations.

\section{Simulations}
\label{sec:sims}

\begin{figure*}
\centering
\includegraphics[width=0.97\textwidth]{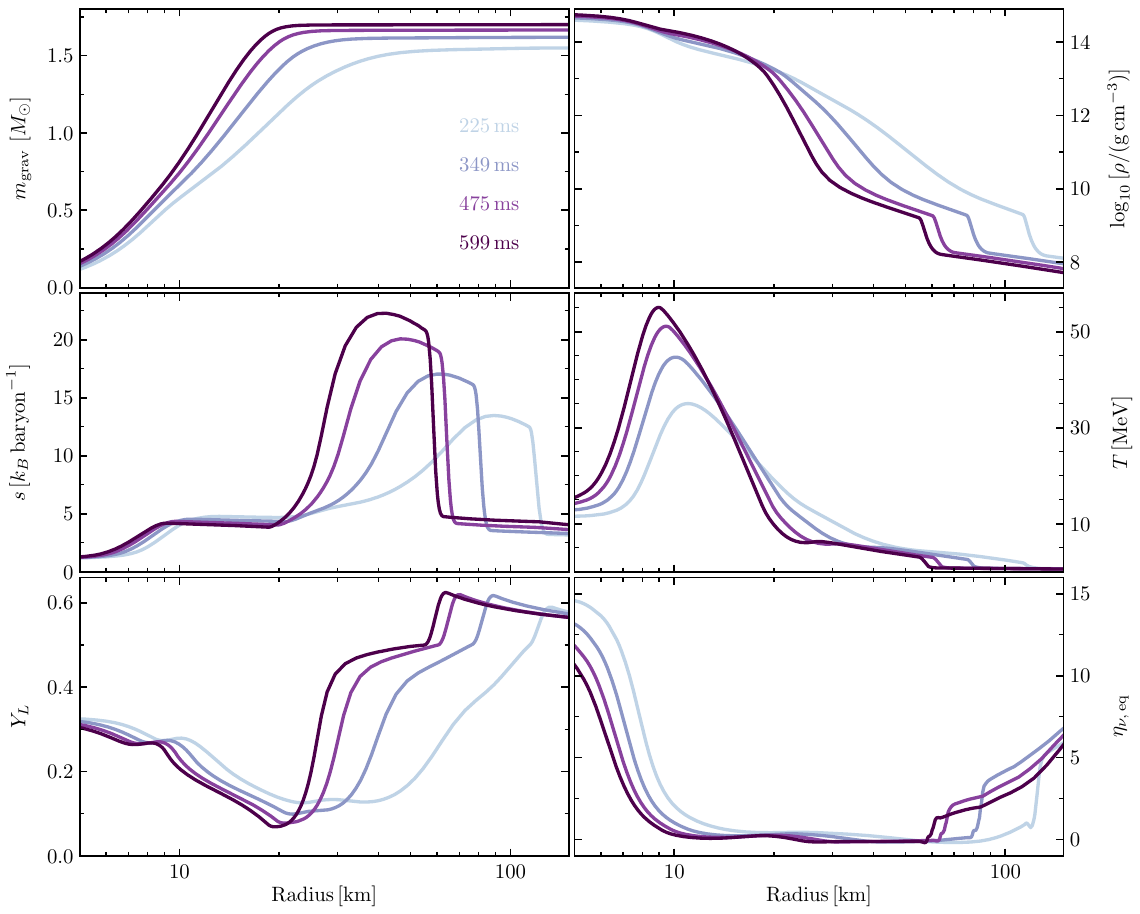}
\caption{Radial profiles of the enclosed gravitational mass $m_{\mathrm{grav}}$ 
(top left-hand panel), logarithmic mass density $\log_{10}\rho$ (top right-hand panel), 
specific entropy $s$ (centre left-hand panel), temperature $T$ (centre right-hand panel), 
lepton fraction $Y_{\mathrm{L}}$ (bottom left-hand panel), and neutrino degeneracy 
parameter $\eta_{\nu,\,\mathrm{eq}}$ (bottom right-hand panel) inside the PNS at 
four times after core bounce. Snapshots at $225$, $349$, $475$, and $599\,\mathrm{ms}$ are shown in light blue, lavender, 
rich lilac, and aubergine, respectively.}\label{fig:6panelstructureplot}
\end{figure*}

We employ a spherically symmetric, fully implicit, general relativistic         
radiation hydrodynamics code, which employs mixing length theory to account for 
the effect of convective energy                 
transport~(\citealt{roberts:2012:thesis,roberts:2012}). Neutrino transport is   
treated through a general-relativistic moment-wise scheme, employing a variable 
Eddington factor approach which retains only the first two moment equations 
and assumes a closure relation between these and higher order moments 
(see~\citealt{mihalas:mihalas:1984}) derived from a formal solution of the 
static relativistic Boltzmann equation~\citep{roberts:2012}. The approach, 
which incorporates inelastic scattering and pair production, treats the 
spectral behaviour of the neutrinos via energy-integrated groups. Three species 
of neutrino are considered, $\nu_{i}\in\{\nu_{e}, \bar{\nu}_{e}, \nu_{x}\}$, 
where $\nu_{x}$ is a characteristic heavy-lepton neutrino employed to encompass 
the effects of neutrinos $\nu_{\mu}$, $\nu_{\tau}$, and their antiparticles. 
Each neutrino population is modelled by a distribution of massless fermions, 
where neutrinos within a given energy bin are distributed as for a Fermi 
blackbody. For each energy group, the effective Planck mean opacity from 
absorption is computed using a 10-point quadrature to calculate the 
group-averaged opacities, corrected for detailed balance. Group-averaged 
opacities from scattering, calculated using a five-point quadrature, are not 
weighted by the local thermal neutrino distribution.

We simulate the core collapse and post-bounce evolution of a $15\,\msun$           
progenitor star from \citet{woosley:weaver:1995} out to $660\,\mathrm{ms}$ 
after core bounce. To close the set of general-relativistic radiation 
hydrodynamic equations, we employ the Lattimer-Swesty equation of state with 
incompressibility parameter $K = 220\,\mathrm{MeV}$, modified as outlined     
in~\citet{lattimer:swesty:1991} and \citet{schneider:etal:2017}. The prescription for       
energy-dependent neutrino transport uses 20 energy groups covering          
$[0,\,200]\,\mathrm{MeV}$. Here, 19 logarithmically spaced energy bins    
span the range $[0,\,80]\,\mathrm{MeV}$, and a final group spans                
$[80,\,200]\,\mathrm{MeV}$. In Fig.~\ref{fig:6panelstructureplot}, we show      
snapshots of the radial profiles of interior PNS quantities at 
$225$, $349$, $475$, and $599\,\mathrm{ms}$ after core bounce.

The entropy in the post-shock region rises from $\sim12\kbpbaryon$ to 
$\sim22\kbpbaryon$ between $225$ and $600\,\mathrm{ms}$ after core 
bounce. Gradients in composition and temperature in the PNS mantle steepen with 
time, driving convective instability there from $\sim200\,\mathrm{ms}$ after 
core bounce through the end of the simulation. The peak temperature, located 
around the inner boundary of the inner convective zone, rises from 
$\sim30\,\mathrm{MeV}$ to more than $50\,\mathrm{MeV}$ between 
$225$ and $660\,\mathrm{ms}$ as the PNS cools and contracts. 
Comparatively, low lepton fractions in the PNS core are accompanied by strongly 
degenerate $\nu_{e}$ neutrinos. In the domain 
$10\mathkm \lesssim r \lesssim 20\mathkm$, convection is driven by negative 
lepton and entropy gradients, with convective transport pushing the unstable 
region towards neutral buoyancy.

\section{Wave generation and energy transport}
\label{sec:wavegen}
Gravity waves are excited by turbulent convection at the interface between 
convective and radiative zones. To establish said boundaries, we use the sign 
of the squared Brunt-V\"{a}is\"{a}l\"{a} (or buoyancy) frequency $N^{2}$ as a 
proxy for the convective stability of a particular region, where 
\begin{equation}                                
N^{2} = -\frac{g}{n_{B}}\left( \frac{d n_{B}}{dr} -                             
\frac{dP}{dr} \left(\frac{\partial n_{B}}{\partial P}\right)_s \right)\,,       
\end{equation}                                                                  
given the local gravitational acceleration $g$ and number density of 
baryons $n_{B}$. We designate radial grid zones convective where $N^{2}<0$, 
and conversely radiative grid zones where $N^{2}>0$. For plotting purposes, 
the scaled buoyancy frequency $N_{p}$ is also an instructive quantity to 
calculate, where  
\begin{equation}                                                                
N_{p} = \mathrm{sign}\left( N^{2} \right) \sqrt{|N^{2}|}\,.                     
\end{equation}

\begin{figure}                                                                  
\centering                                                                      
\includegraphics[width=0.97\columnwidth]{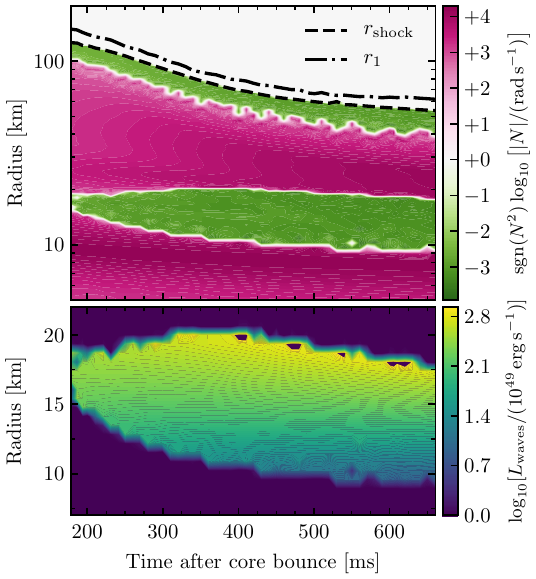}          
\caption{(Top panel) Colormap showing the time evolution of the scaled          
Brunt-V\"{a}is\"{a}l\"{a} frequency $N_{p}=\mathrm{sign}(N^{2})\sqrt{|N^{2}|}$ 
(units of $10^{3}\radpers$) as a function of radius $r$ between 
$200\,\mathrm{ms}$ and $660\,\mathrm{ms}$ after core bounce. The shock front 
is bounded by inner radius $\rshock$ (shown with dashed black line) and outer 
radius $r_{1}$ (shown with dot-dashed black line). Convectively unstable 
regions ($N_{p}<0$) are shown in green, while radiative regions ($N_{p}>0$, 
where gravity waves can propagate), are shown in pink. (Bottom panel) Temporal 
evolution of the characteristic wave luminosity $\lwave = \Mcon\lcon$ generated 
by turbulent convection in the inner convective region.}                                     
\label{fig:kippenhahn}                                                          
\end{figure}

In the top panel of Fig.~\ref{fig:kippenhahn}, we show the temporal evolution   
of $N_{p}$ below the shock. We see the emergence of a convectively unstable     
region around $20\mathkm$ a little before $200\,\mathrm{ms}$ after bounce, which 
quickly develops into a convective layer between $10$ and $20\mathkm$ 
that persists through the end of our simulation. We note that, in the region of 
interest (the outer boundary of the inner convective region), $N_{p}$ is of 
order $\sim10^{3}\radpers$.

The emitted flux in gravity waves $\lwave$ is a small fraction of the 
convective energy flux $\lcon$. While there is some uncertainty in the wave 
spectrum and energy flux, both 
analytic~\citep{goldreich:kumar:1990,lecoanet:quataert:2013} and numerical      
work~\citep{couston:etal:2018} suggest the wave energy flux is approximately    
\begin{equation}                                                                
\label{lwaves}                                                                  
    \lwave \sim \Mcon\lcon\,,                           
\end{equation}                                                                  
where the convective Mach number $\Mcon = \vcon/c_{s}$ can be calculated using 
the characteristic convective velocity                                              
\begin{equation}                                                                
\vcon = \left[ \frac{\lcon}{4\pi r^{2}\rho} \right]^{1/3}\,.                    
\end{equation}                                                                  
In practice, we can use the total outgoing neutrino luminosity $\lnu$ to        
estimate $\vcon$ as, in the optically thick inner PNS where convection is       
efficient and carries almost all of the energy flux, $\lcon \sim \lnu$. 
For the purposes of this study, we employ $\lwave$ as 
estimated at the upper edge of the inner convective region, where we expect 
the waves to emerge. In the bottom panel of Fig.~\ref{fig:kippenhahn}, we 
show the evolution of $\lwave$ across the inner convective zone with 
time. At the upper edge, we see wave luminosities in excess of 
$10^{51}\ergpers$ sustained from around $200\,\mathrm{ms}$ after core bounce 
through the end of our simulation.



\begin{figure}                                                                  
\centering                                                                      
\includegraphics[width=0.97\columnwidth]{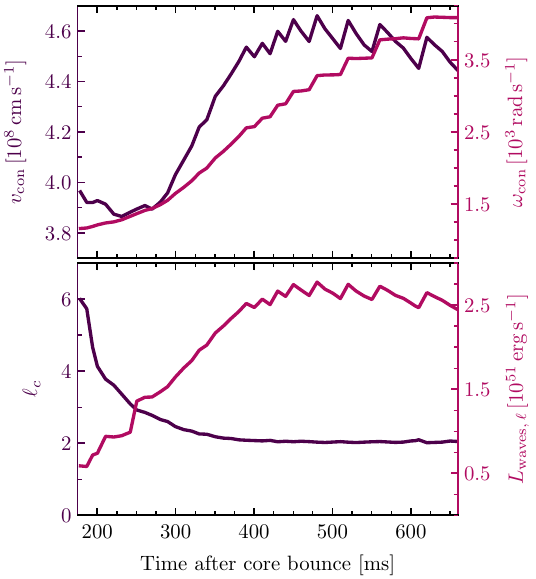}
\caption{(Top panel) The evolution of the convective velocity $\vcon$ 
(left axis, deep purple line) and convective turnover frequency $\wcon$ 
(right axis, deep pink line) at the upper edge of the inner convective region 
with time. (Bottom panel) The evolution of the maximum angular wave mode 
excited $\ellc$ (left axis, deep purple line) and emitted wave luminosity per 
excited angular mode $\ell$, $\lwavel$ (right axis, deep pink line) near the 
upper edge of the inner convective region are shown with time.}
\label{fig:wave-props}                                                          
\end{figure}

We consider waves generated by convective motion within a pressure scale        
height $H_{p}$ of the outer boundary of the inner convective region (at radius 
$r = \rcon$), where                                      
\begin{equation}                                                                
    H_{p} = \frac{P}{g(\rho + P/c^{2})}\,.                                      
\end{equation}                                                                  
While the frequency and angular wavenumber spectra of waves generated in this   
scenario are decidedly 
uncertain~\citep{lecoanet:quataert:2013,rogers:etal:2013,alvan:etal:2014,couston:etal:2018},
most literature agrees that the wave power drops significantly at frequencies   
above the local convective turnover frequency $\wcon$, which we define as       
\begin{equation}                                                                
\wcon = \frac{\pi \vcon}{2H_{p}}\,.                                             
\end{equation}                                                                  
We adopt a flat spectrum across angular modes $\ell \in [1,\ldots,\ellc]$,      
where $\ellc = \rcon/\drcon$ is determined by the physical width of the 
convection zone, $\drcon$. This choice reflects the fact that thin convective 
zones have smaller eddies that excite waves with smaller horizontal wavelength 
and thus larger angular wavenumber $\ell$. The emitted wave flux per mode 
$\lwavel$ is then just                   
\begin{equation}                                                                
\lwavel = \frac{\lwave}{\ellc}\,.                                                         
\end{equation}

\begin{figure*}                                                                 
    \centering                                                                  
    \includegraphics[width=0.97\textwidth]{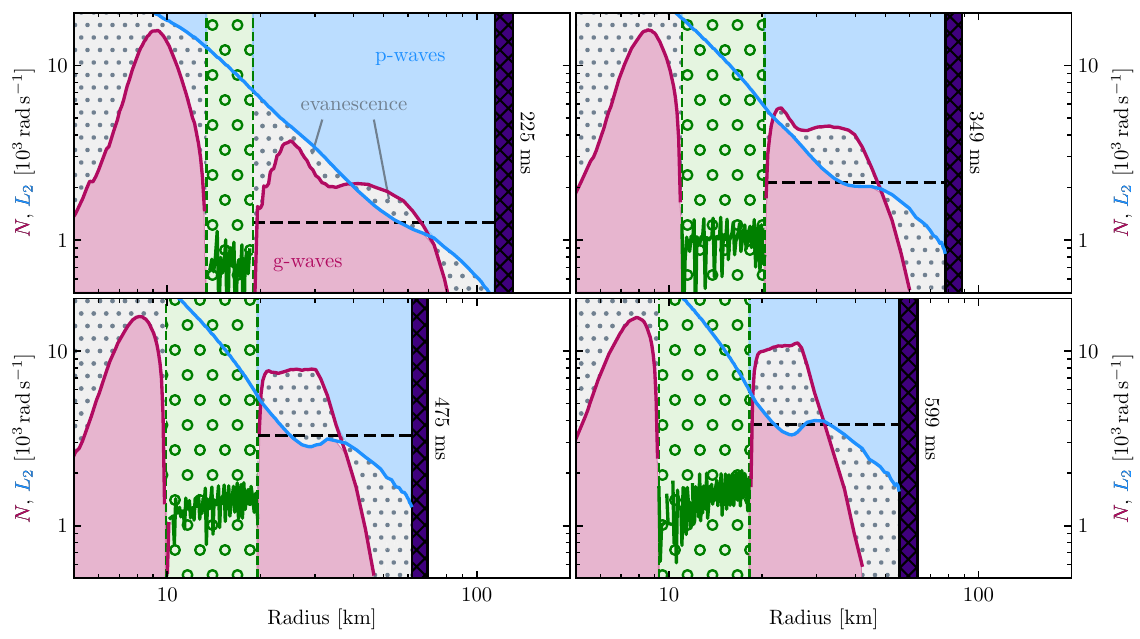}
    \caption{Propagation diagrams for $\ell=2$ gravitoacoustic waves at        
    $225\,\mathrm{ms}$ (top left-hand panel), $349\,\mathrm{ms}$ (top right-hand panel),  
    $475\,\mathrm{ms}$ (bottom left-hand panel), and 
    $599\,\mathrm{ms}$ (bottom right-hand panel) after core bounce. In each panel,   
    the inner convective region is shown with green shading with circle hatch, 
    with the magnitude of the scaled buoyancy frequency $|N_{p}|$ overlaid. The 
    $\ell=2$ Lamb frequency $L_{2}$ is shown in blue and, in radiative regions, 
    the scaled buoyancy frequency $N_{p}$ is shown in deep pink. Regions where  
    gravity waves of angular frequency $\omega$ can propagate                   
    (i.e., where $\omega<N,\,L_{2}$) are shaded in pink, while regions in which 
    acoustic waves can propagate (i.e., where $\omega>N,\,L_{2}$) are shaded in 
    blue. Waves cannot propagate in evanescent regions 
    (where $N<\omega<L_{2}$ or $N>\omega>L_{2}$), shown with grey shading and   
    dotted hatch. The physical extent of the shock is indicated with dark       
    purple shading and crossed hatch. For waves at frequency $\wcon$, the       
    propagation track is shown with a thick dashed black line.}                 
    \label{fig:propagation}                                                     
\end{figure*}

In Fig.~\ref{fig:wave-props}, we show the evolution of the convective           
velocity $\vcon$, convective turnover frequency $\wcon$, maximum angular        
wavenumber excited $\ellc$, and the wave flux per angular mode $\lwavel$ with   
time. We find the convective velocity remains roughly constant around           
$\vcon\sim(4-4.5)\times10^{8}\cmpers$, while the convective turnover frequency 
steadily increases by a factor of roughly four from $\wcon\sim10^{3}\radpers$ 
to $\wcon\sim4\times10^{3}\radpers$ between $200\,\mathrm{ms}$ post-bounce 
through the end of our simulation. At early times, when the physical width of 
the inner convective zone is small, modes up to $\ellc\sim(5-6)$ are excited, 
though $\ellc\lesssim 3$ by $\sim \! 250\,\mathrm{ms}$ post-bounce. From 
$\sim \! 250\,\mathrm{ms}$ through the end of our simulation, the emitted wave 
luminosity per angular mode exceeds $10^{51}\,\ergpers$.

After waves are excited, their propagation within the PNS 
is largely governed by the dispersion relation for gravitoacoustic waves of 
angular frequency $\omega$. In the WKB approximation (see, e.g., Eq. S7 
in~\citep{fuller:etal:2015:b}), the radial wavenumber $k_{r,\,\ell}$ for waves 
of angular mode $\ell$ is 
\begin{equation}
\label{krl}
k^{2}_{r,\,\ell} = \frac{(|N|^{2} -
\omega^{2})(L_{\ell}^{2}-\omega^{2})}{\omega^{2}c_{s}^{2}}\,,                   
\end{equation}                                                                  
where Lamb frequencies $L_{\ell}$ are defined                                     
\begin{equation}                                                                
L_{\ell}^{2} = \frac{\ell(\ell+1)c_{s}^{2}}{r^{2}}\,,                           
\end{equation}                                                                  
given adiabatic sound speed $c_{s}^{2} = \Gamma_{1} P/\rho$, where                
$\Gamma_{1}=(\partial \log P/\partial \log \rho )_{s}$ is the adiabatic index.  
Where $\omega<N,L_\ell$, the waves propagate as gravity waves, while they       
propagate as acoustic waves where $\omega > N,L_\ell$. Where 
$N<\omega<L_{\ell}$ and $L_{\ell}<\omega<N$, waves cannot propagate and are     
evanescent. In Fig.~\ref{fig:propagation}, we show propagation diagrams for     
$\ell=2$ waves at $225\,\mathrm{ms}$, $349\,\mathrm{ms}$, $475\,\mathrm{ms}$,   
and $599\,\mathrm{ms}$ after core bounce. From Fig.~\ref{fig:propagation}, we see that the width of the evanescent region 
is highly sensitive to the wave frequency. For $\ell=2$ waves, the propagation  
track along $\omega=\wcon$ skims the bottom of a large evanescent region in     
frequency space. After one pass, a fraction $T_{\ell}^{2}$ of the incident wave 
flux is transmitted through the evanescent region (bounded by gravitoevanescent
radius $r_{\mathrm{grav-ev}}$ and acoustic-evanescent radius                    
$r_{\mathrm{ac-ev}}$), where transmission coefficients $T_{\ell}$ are           
\begin{equation}                                                                
T_{\ell} = \exp\left[-\int_{r_{\mathrm{grav-ev}}}^{r_{\mathrm{ac-ev}}}          
dr\,|k_{\ell}|\right]\,.                                                        
\end{equation}                                                                  
While, in principle, more wave energy could leak through the boundary 
following multiple reflections within the PNS, we adopt 
$\ltransl = \lwavel T_{\ell}^{2}$ as a conservative lower limit on the wave 
flux in angular mode $\ell$ entering the outer PNS.

\begin{figure*}                                                                 
\centering                                                                      
\includegraphics[width=0.97\textwidth]{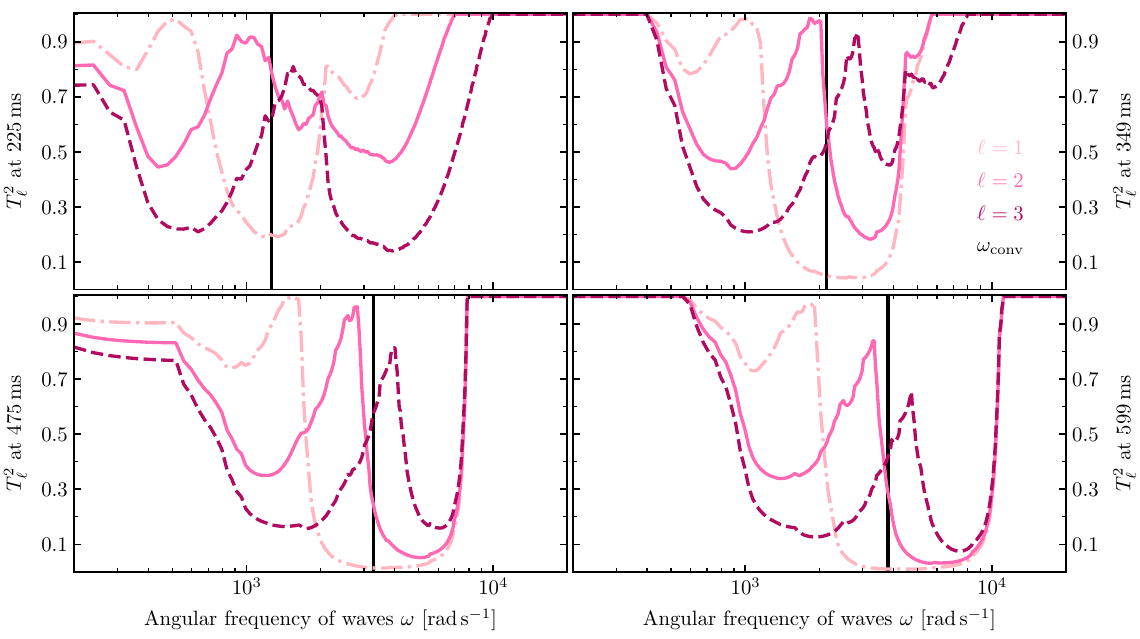}      
\caption{(Top left-hand panel) Snapshots of the squared transmission coefficients    
$T^{2}_{\ell}$ for $\ell=1$~(dash-dotted light pink line),                      
$\ell=2$~(solid hot pink line), and $\ell=3$~(dashed deep pink line) waves      
as a function of angular wave frequency $\omega$ at $225\,\mathrm{ms}$          
post-bounce. The convective turnover frequency $\wcon$, at which we assume      
waves are excited, is indicated with a solid black line. Similar snapshots at   
$349\,\mathrm{ms}$ (top right-hand panel), $475\,\mathrm{ms}$ (bottom left-hand panel),   
and $599\,\mathrm{ms}$ (bottom right-hand panel) post-bounce are shown.}             
\label{fig:transmission}                                                        
\end{figure*}

In Fig.~\ref{fig:transmission}, we show snapshots of $T_{\ell}^{2}$, the        
fraction of wave energy transmitted out to the post-shock region, for           
$\ell=1-3$ waves as a function of angular frequency at $225\,\mathrm{ms}$,      
$349\,\mathrm{ms}$, $475\,\mathrm{ms}$, and $599\,\mathrm{m}$ after core bounce.
We see that uncertainty in $\wcon$ translates to uncertainties of order unity   
in the wave flux transmitted through the evanescent region. In reality, a       
spectrum of waves extending above and below $\omega_{\rm con}$ will be 
generated, and the energy transmitted into acoustic waves of wavenumber $\ell$ 
will be $\int d \omega \, T_{\ell}^{2} \, (d\lwavel/d\omega)$, where 
$(d\lwavel/d\omega)$ is the energy flux in waves generated per unit frequency 
$\omega$. As the wave spectrum is uncertain, this integral   
is uncertain, but Fig.~\ref{fig:transmission} demonstrates that we expect       
transmission fractions of order $T_{\ell}^{2} \sim 0.3$, dependent on the       
excited wave spectrum and time after core bounce.

\begin{figure}                                                                  
\centering                                                                      
\includegraphics[width=0.97\columnwidth]{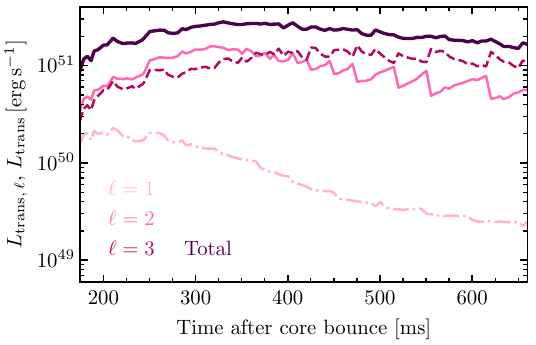}                     
\caption{Temporal evolution of the transmitted wave power 
$L_{\mathrm{trans},\ell}$ for $\ell=1$~(light pink dash-dotted line), 
$\ell=2$~(solid hot pink line), and $\ell=3$~(dashed deep pink line) waves 
in the post-shock region following core bounce. The total transmitted wave 
power $\ltrans$, summed over all excited wave modes, is overlaid with a thick 
deep purple line.}
\label{fig:heating}                                                             
\end{figure}

Following their transmission out the upper edge of the evanescent region,       
acoustic waves can travel almost unimpeded out to the shock. The wave energy    
transport rate (i.e., ``transmission rate") beyond the last evanescent region        
encountered is given by                                                         
\begin{align}                                                                   
\ltrans = \sum_{\ell=1}^{\ellc}\,\ltransl\,,\\
\ltransl = \lwavel T_{\ell}^{2}\,.
\end{align}                                                                     
We show the temporal evolution of the transmitted wave power for $\ell=1-3$ 
angular modes and the summed total in the post-shock region in 
Fig.~\ref{fig:heating}. Once the inner PNS convection zone forms and begins 
generating waves, we find net transmitted wave power $\ltrans$ in excess of 
$\sim \! 10^{51}\ergpers$ through the end of our simulation. The power 
contributed by $\ell=1$ waves drops significantly with time, with transmitted 
power in $\ell=2$ and $\ell=3$ waves responsible for most of the net heating in 
the post-shock region. As illustrated by 
Fig.~\ref{fig:transmission}, a more accurate calculation that integrates over 
the wave spectrum may change the transmission profile. For example, at 
$349\,\mathrm{ms}$ post-bounce (as shown in the top right-hand panel of 
Fig.~\ref{fig:transmission}), we see that lower frequency $\ell=1$ waves and 
higher frequency $\ell=3$ waves would be transmitted more easily. While 
this might yield slightly different net wave power transmission rates, we 
expect our results to be accurate within a factor of order unity.

\section{Wave damping and non-linear effects}
\label{sec:damping-and-non-lin}
Until now, our calculations have implicitly assumed that waves generated by PNS 
convection propagate according to linear perturbation theory, and we have       
ignored sources of damping. We now address the validity of these assumptions,   
and work to quantify the effect of wave damping and non-linearity.

\subsection{Neutrino damping}
\label{subsec:nudamp}
As discussed in~\citet{weinberg:quataert:2008}, gravity wave attenuation        
through radiative losses is dominated by increased neutrino emission from       
regions of wave compression. Here, we consider the neutrino damping of gravity  
waves emitted from the inner convective zone as they propagate from the PNS     
core out to the stalled shock.

The energy loss rate from neutrino damping, $\dot{\varepsilon}_{\nu}$, can be 
calculated as
\begin{align}
    \dot{\varepsilon}_{\nu} &= \delta T\, \delta \left( \frac{\mathrm{d}s}{\mathrm{d}t} \right)_{\nu}\,,
\end{align}
where $\delta$ denotes the Lagrangian variation of the following term, and 
$\mathrm{d}s/\mathrm{d}t$ is the rate of change of specific entropy due to neutrino 
dissipation. While accurately calculating $\delta (\mathrm{d}s/\mathrm{d}t)_{\nu}$ in the neutrino
decoupling region requires a full solution of the neutrino transport equations, 
we can consider in turn the limits of optically thick damping (where wavelength 
$\lambda_{r,\ell} \sim 2\pi/|k_{r,\ell}|$ exceeds the neutrino 
mean free path $\langle d_{\nu_{i}} \rangle$) and optically thin damping 
($\lambda_{r,\ell} < \langle d_{\nu_{i}} \rangle$) to simplify matters 
considerably.

In the optically thick limit, neutrino losses are dominated by the 
diffusion of $\mu$ and $\tau$ neutrinos, for which the mean free path is 
determined pre-dominated by neutral current scattering on nucleons. It can 
be shown (see Section~\ref{app:thick}) that 
\begin{align}
    \dot{\varepsilon}_{\nu,\mathrm{thick}} &\approx 
    \frac{c\, (k_B m_{e}c^{2})^{2}}{9\,(\hbar c)^{3}\,\sigma_{0}} 
    \frac{|k_{r,\ell}|^{2}\,\delta T^{2}}{n_{B}^{2}\,C(Y_{e})}\,,
\end{align}
where $\sigma_{0} = 1.7 \times 10^{-44} \, \textrm{cm}^2$ is the 
fiducial weak interaction cross section, and
\begin{align}
    C(Y_{e}) &=  \frac{1 + 5g_{A}^{2}}{6} \frac{1-Y_{e}}{4} + 
    \frac{(C^{'}_{V} - 1)^{2} + 5g_{A}^{2} (C^{'}_{A} - 1)^{2} }{6} Y_{e}\,,
\end{align}
encompasses compositional effects. Here, 
$g_{A}\approx-1.27$ is the charged current axial coupling constant, while 
$C^{'}_{A}=0.5$ and $C^{'}_{V}\approx0.96$ are the neutral current axial and 
vector coupling constants, respectively.

Conversely in the optically thin limit, neutrino losses are 
dominated by emission and subsequent free-streaming of $\nu_{e}$ and 
$\bar{\nu}_{e}$ neutrinos, which is driven by charged-current capture on 
nucleons. In this scenario, the energy loss rate in neutrinos can be 
shown (see Section~\ref{app:thin}) to take the form;
\begin{align}
\dot{\varepsilon}_{\nu_{e}+\bar{\nu}_{e},\,\mathrm{thin}} &\approx 
    \frac{3c\sigma_{0}k_B^6}{4\pi^{2}(\hbar c)^{3}(m_{e}c^{2})^{2}}\,C_{\nu_{e}+\bar{\nu}_{e},\,\mathrm{thin}}\,T^{4}\,\delta T^{2}\,,    
\end{align}
where, we have defined 
\begin{align}
    C_{\nu_{e}+\bar{\nu}_{e},\,\mathrm{thin}} &= (1+3g_{A}^{2}) \notag\\ 
    &\times \bigg[ Y_{e} \int_{\Delta_{np}/T}^{\infty} dx\,x^{5} \left(1 - \frac{\Delta_{np}}{xT}\right)^{2} \frac{1 + e^{-x - \eta_{\nu_{e},\mathrm{eq}}}}{1+ e^{x + \eta_{\nu_{e},\mathrm{eq}}}} \notag\\
    +  &(1- Y_{e}) \int_{0}^{\infty} dx\,x^{5} \left(1 + \frac{\Delta_{np}}{xT}\right)^{2} \frac{1 + e^{-x + \eta_{\nu_{e},\mathrm{eq}}}}{1+ e^{x - \eta_{\nu_{e},\mathrm{eq}}}} \bigg]\,,
\end{align}
to encompass coupling constants and compositional effects.

Requiring continuity in the energy damping rates between the optically thick 
and optically thin limits, we may define some critical wavenumber 
$k_{\mathrm{crit}}$ such that 
\begin{align}
    \dot{\varepsilon}_{\nu,\,\mathrm{thick}} \left( k = k_{\mathrm{crit}}\right) = 
    \dot{\varepsilon}_{\nu,\,\mathrm{thin}} \left( k = k_{\mathrm{crit}}\right)\,.
\end{align}
Using the expressions derived above, we find
\begin{align}
    k_{\mathrm{crit}} &= \frac{3\,\sigma_{0}\, n_{B}\, (k_B T)^{2} }{ 2 \pi (m_{e}c^{2})^{2} }  
    \sqrt{3\, C(Y_{e})\, C_{\nu_{e}+\bar{\nu}_{e},\,\mathrm{thin}}}\,.
\end{align}
Given this threshold, we treat wave damping via neutrino dissipation in the 
optically thick limit where $k<k_{\mathrm{crit}}$, and conversely in the 
optically thin limit where $k>k_{\mathrm{crit}}$.

For waves with angular frequency $\omega$, the damping frequency 
$\omega_{\mathrm{damp}}$ is defined
\begin{align}
    \omega_{\mathrm{damp}} \equiv \frac{\dot{\varepsilon}_{\nu}}{\varepsilon_{\mathrm{wave}}}\,,
\end{align}
where $\varepsilon_{\mathrm{wave}}\sim \omega^{2} |\xi|^{2}$ is the wave energy 
per unit mass, and $|\xi|$ is the magnitude of the wave displacement.

The Lagrangian temperature perturbation produced under passage of the wave is given 
by
\begin{align}
    \frac{\delta T}{T} \sim |k_{r,\ell}|\,\xi_{r}\,,
\end{align}
where $\xi_{r}$ is the radial component of the wave displacement. For gravity waves, 
$\xi_{r} \sim (\omega/|N|)\,|\xi|$, while  
$\xi_{r} \sim |\xi|$ for acoustic waves. The damping frequency may then be expressed more generally as
\begin{align}
    \omega_{\mathrm{damp}} &\approx \frac{|k_{r,\ell}|^{2}}{\omega^{2}}\,
    \mathrm{min}\left[1,\,\frac{\omega^{2}}{|N|^{2}} \right] \notag\\
    &\times 
    \begin{cases}
        \frac{c\,(k_B m_{e}c^{2})^{2}}{9\,(\hbar c)^{3}\sigma_{0}} \frac{|k_{r,\ell}|^{2}}{n_{B}^{2}\,C(Y_{e})}\,T^{2} & k \leq k_{\mathrm{crit}}\,,\\
        \frac{3 c \sigma_{0} k_B^6}{4\pi^{2} (\hbar c)^{3} (m_{e}c^{2})^{2} }\,C_{\nu_{e}+\bar{\nu}_{e},\mathrm{thin}}\, T^{6} & k > k_{\mathrm{crit}}\,.
    \end{cases}
\end{align}

\begin{figure*}                                                                 
\centering                                                                      
\includegraphics[width=0.97\textwidth]{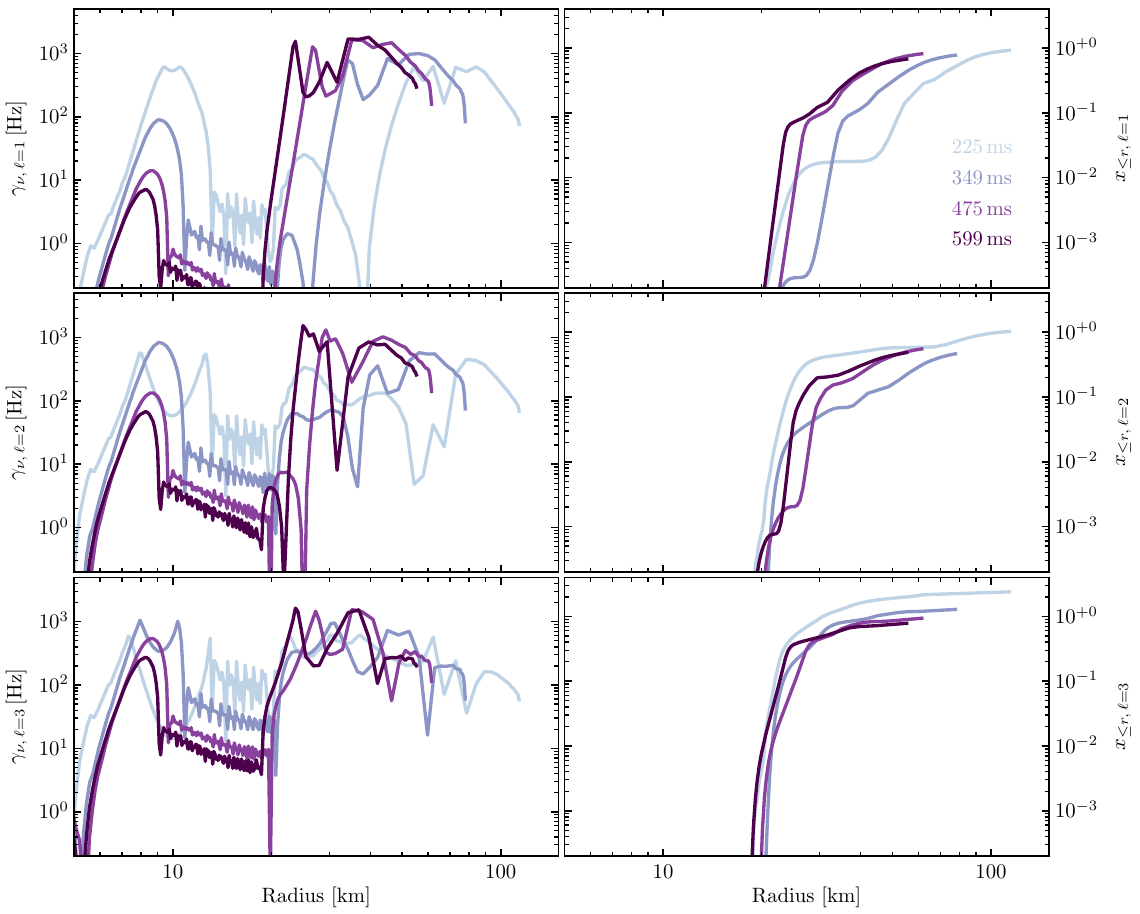}                     
\caption{Snapshots of the local damping rate from neutrino diffusion            
$\gamma_{\nu}$ (left column), and the cumulative damping incurred by waves      
$x(\leq r)$ while propagating out from the outer edge of the inner PNS 
convective region $\rcon$ to radius $r$ (right column). In each panel, radial 
traces are shown at $225\,\mathrm{ms}$ (in light blue), $349\,\mathrm{ms}$ 
(in lavender), $475\,\mathrm{ms}$ (in rich lilac), and $599\,\mathrm{ms}$ 
(in aubergine) for $\ell=1$ (top panels), $\ell=2$ (centre panels), and 
$\ell=3$ (bottom panels) waves.}
\label{fig:gammanu}                                                             
\end{figure*}

By incorporating the neutrino energy loss term into the energy evolution 
equation, one can show that it causes the wave frequency $\omega$ to become 
imaginary with damping rate $\gamma_{\nu}$. In the quasi-adiabatic limit 
where $\gamma_{\nu}\ll\omega$, the wave amplitude damping rate is 
\begin{align}
    \gamma_{\nu} \approx \frac{1}{2}\omega_{\mathrm{damp}}\,,
\end{align}
while in the isothermal limit (where heat diffusion is fast), we find
\begin{align}
    \gamma_{\nu} \approx \frac{1}{2}\frac{\omega^{2}}{\omega_{\mathrm{damp}}}\,.
\end{align}
Generally speaking, the wave damping rate may be written
\begin{align}
    \gamma_{\nu} &\approx \frac{\omega_{\mathrm{damp}}}{2} \times 
    \mathrm{min}\bigg[1,\,\frac{\omega^{2}}{\omega^{2}_{\mathrm{damp}}}  \bigg]\,.
\end{align}
In the left column of Fig.~\ref{fig:gammanu}, we show radial profiles of the    
local neutrino damping rate $\gamma_{\nu,\ell}$ for $\ell=1-3$ waves at 
$225\,\mathrm{ms}$, $349\,\mathrm{ms}$, $475\,\mathrm{ms}$, and 
$599\,\mathrm{ms}$ after core bounce.

The wave flux, propagating from the outer boundary of the inner convective 
region ($\rcon$) out towards the shock, is attenuated by a factor 
$e^{-x(\leq r)}$ after travelling to radius $r$, where $x(\leq r)$ is defined
\begin{equation}    
x(\leq r) = \int_{\rcon}^{r} \frac{dr^{'}\,2 \gamma_{\nu}(r^{'})}{v_{\mathrm{gr}}}\,. 
\end{equation} 
Here, $v_{\mathrm{gr}} = |\omega/k_{r}|$ is the waves' group velocity. In the   
right column of Fig.~\ref{fig:gammanu}, we show radial snapshots of the         
cumulative damping experienced, $x(\leq r)$, for $\ell=1-3$ waves at            
$225$, $349$, $475$, and                 
$599\,\mathrm{ms}$ post-bounce.

We see that damping rates are typically largest in the outer PNS for all $\ell = 1-3$ waves. This makes sense, as neutrino cooling rates are largest in the outer PNS, and the neutrino damping operates in the optically thin limit. While optically thick wave damping in the inner core can be important at early times, this does not affect the outgoing waves of interest here. Neutrino damping becomes less important near the shock because the neutrino emissivities become smaller. Although outgoing waves can be rapidly damped in the outer PNS, their large group velocities prevent their attenuation factors $x(\leq r)$ from greatly exceeding unity.

\begin{figure}
\centering
\includegraphics[width=0.97\columnwidth]{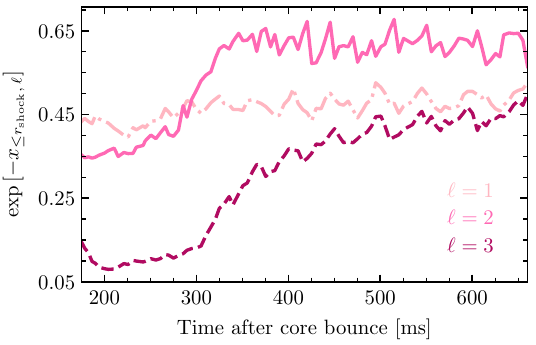}                    
\caption{Temporal evolution of the fraction of wave luminosity not attenuated   
through neutrino damping (as measured just below the shock) with post-bounce    
time for $\ell=1$ (dash-dotted light pink line), $\ell=2$ (solid hot pink line),
and $\ell=3$ (dashed dark pink line) waves.}
\label{fig:nudamp}
\end{figure}

The total fraction of unattenuated wave energy (i.e., the fraction of wave 
energy not lost to neutrino damping in travelling from the PNS out to the shock) 
is shown as a function of time in Fig.~\ref{fig:nudamp} for $\ell = 1-3$ waves. 
We expect neutrino damping to be somewhat important, with a moderate fraction of   
the wave energy lost within and below the gain region through neutrino emission. 
We discuss the ramifications of this, in addition to other potential concerns, in          
Section~\ref{sec:discussion}.

\subsection{Non-linear effects}
\label{subsec:nonlin}
Waves excited by convection will continue to propagate inside the PNS until 
their energy is dissipated as heat. In the absence of damping, the wave 
displacement $\mathbf{\xi}$ (dependent on the total wave energy) grows as the 
waves propagate outwards into regions with lower density, and non-linear 
effects may quickly become non-negligible. A useful measure of non-linearity    
for both gravity waves and acoustic waves is the dimensionless quantity         
$|\mathbf{k}\cdot\mathbf{\xi}| \sim |k_{r}\xi_{r}|$. Previous work 
(e.g.,~\citealt{barker:ogilvie:2011}) has shown that as 
$|k_{r}\xi_{r}|\rightarrow1$, gravity waves overturn via stratification and     
break, losing energy via Kelvin-Helmholtz instabilities. As acoustic waves enter
the non-linear regime, energy is lost as the waves self-shock and dissipate.

In the limit of no damping, it can be shown that the wave amplitude $|\xi|$     
associated with energy flux $\ltrans$ is approximately~\citep{fuller:ro:2018}
\begin{equation}                                                                
|\xi| \approx \left[ \frac{\ltrans}{4\pi \rho r^{2}\omega^{2}v_{\mathrm{gr}}} \right]^{1/2}\,.
\end{equation}                                                                  
Since gravity waves propagate in the inner PNS and mantle before coupling to    
acoustic waves near the post-shock gain region, we consider both these cases in 
turn.

For gravity waves, the wave amplitude is 
$|\xi| \sim (|k_{r}/k_{\perp}|)\xi_{r}$, and the group velocity is 
$v_{\mathrm{gr}}\sim\omega^{2}r/(N\sqrt{\ell(\ell+1)})$. For low-$\ell$ gravity 
waves, the appropriate non-linearity measure thus follows
\begin{equation}                                                                
\left|k_{r}\,\xi_{r}\right|_{\mathrm{grav},\,\ell} \approx                        
\left[ \frac{\ltransl\,N\left(\sqrt{\ell(\ell+1)}\right)^{3} }{ 4\pi\rho r^{5}\omega^{4} }  \right]^{1/2}\,.
\end{equation}                                                                  
For acoustic waves, the wave amplitude is $|\xi| \sim \xi_{r}$, and the group 
velocity is $v_{\mathrm{gr}} = c_{s}$. It follows that, for low-$\ell$ acoustic 
waves,       
\begin{equation}
\left|k_{r}\xi_{r}\right|_{\mathrm{ac},\,\ell} \approx \left[ \frac{\ltransl}{4\pi\rho r^{2} c_{s}^{3}} \right]^{1/2}\,.
\end{equation}

\begin{figure*}                                                                 
\centering                                                                      
\includegraphics[width=0.92\textwidth]{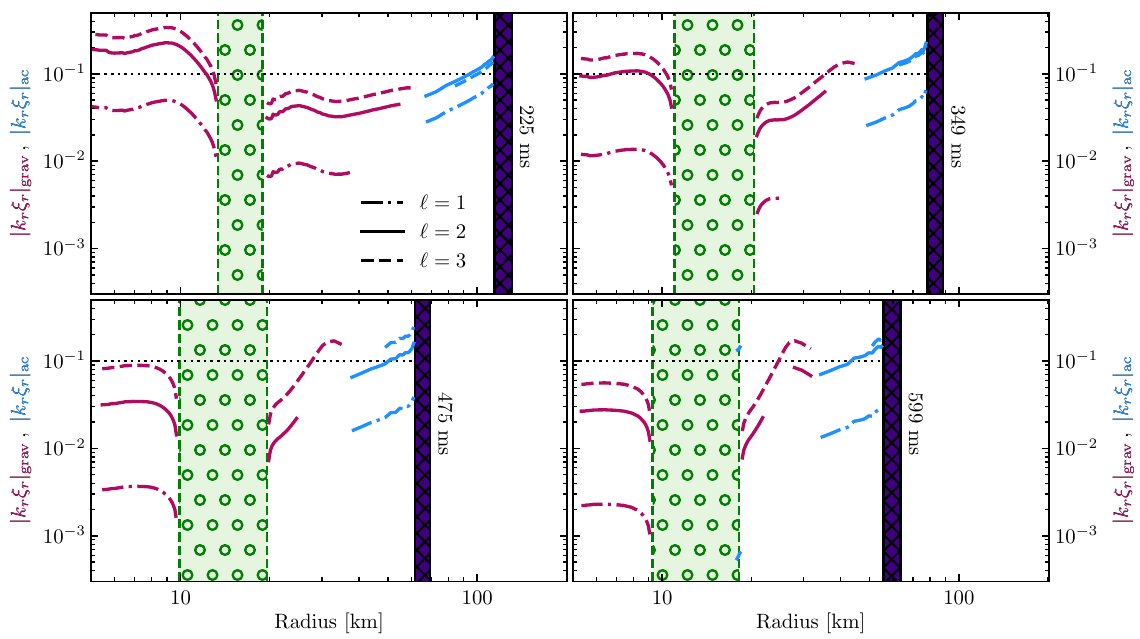} 
\caption{Radial profiles of dimensionless quantity $|k_{r}\,\xi_{r}|_{\ell}$ for
$\ell=1$ (dash-dotted line), $\ell=2$ (solid line), and $\ell=3$ (dashed line)  
waves at $225\,\mathrm{ms}$ (top left-hand panel), $349\,\mathrm{ms}$ (top right-hand 
panel), $475\,\mathrm{ms}$ (bottom left-hand panel), and $599\,\mathrm{ms}$ (bottom 
right-hand panel) after core bounce. Lines are shown in deep pink where gravity waves 
propagate ($\wcon<N,\,L_{\ell}$), and shown in blue where acoustic waves travel 
($\wcon>N,\,L_{\ell}$). Regions in which the waves are evanescent (i.e., where 
$N<\wcon<L_{\ell}$ or $L_{\ell}<\wcon<N$) are left blank. For each snapshot, the 
inner convective region is shown in green with circled hatch, while the shock is 
shown in deep purple with crossed hatch. In each panel, we overlay dotted lines 
at $|k_{r}\,\xi_{r}|_{\ell} = 0.1$, beyond which non-linear effects may 
be important for the wave dynamics.}                                            
\label{fig:nonlin}                                                              
\end{figure*}

In Fig.~\ref{fig:nonlin}, we show radial snapshots of $|k_{r}\xi_{r}|_{\ell}$ 
for $\ell=1-3$ waves at $225\,\mathrm{ms}$, $349\,\mathrm{ms}$, 
$475\,\mathrm{ms}$, and $599\,\mathrm{ms}$ after core bounce. Throughout the 
PNS for the duration of the simulation, $|k_{r}\,\xi_{r}|_{\ell} \ll 1$ for 
$\ell=1$ waves. On approach to the gain region, 
$|k_{r}\,\xi_{r}|_{\ell} \sim 0.1$ for $\ell=2$ and $\ell=3$ waves. In the PNS 
core at early times, non-linear effects may be important for gravity waves. This 
suggests that waves propagating through the inner core are moderately 
non-linear, and may lose some fraction of their energy to non-linear
dissipative effects. As we don't take their energy into account when calculating 
$\ltrans$, such dissipation will not affect our main result, but could limit 
the contribution of such waves to any excess flux above $\ltrans$. The main 
result is that no waves are in the strongly non-linear regime 
($|k_{r}\xi_{r}|_{\ell} > 1)$, so we do not expect rapid wave damping due to 
breaking or shock formation. We do find moderately non-linear amplitudes 
($|k_{r} \xi_{r}|_{\ell} \gtrsim 0.1)$ just below the shock, which could cause 
some wave damping there. When a successful explosion develops, the shock moves 
out to lower densities, and it is much more likely that acoustic waves will 
become strongly non-linear, forming weak shocks and depositing their energy as 
heat within the PNS wind.

It is also important to address the possibility of non-linear three-mode        
coupling and whether this can quench the wave energy transport, as discussed by 
\cite{weinberg:quataert:2008}. The primary waves excited by convection are      
low-order (0-2 radial nodes, $l=1-3$) gravity waves in the outer PNS, whereas   
the acoustic mechanism involves low-order gravity modes in the inner PNS. Due   
to neutrino damping in the outer PNS, the outgoing convectively excited waves   
have low quality factors $Q \sim 3$ (see Fig.~\ref{fig:nudamp}), and there are  
few daughter modes with frequencies 
$\omega_{\mathrm{d}} \sim \omega_{\mathrm{wave}}/2$   
with which to resonantly couple. Mode coupling likely occurs in the             
non-resonant limit given by equation 2 of \cite{weinberg:quataert:2008}, and    
hence we expect saturation energies of 
$E_{\mathrm{sat}}\sim 10^{49}\,\mathrm{erg}$. In contrast to lower frequency 
modes trapped in the inner core, the gravity waves we consider traverse the 
outer PNS in a wave crossing time $t_{\mathrm{cross}} \sim5\,{\mathrm{ms}}$. 
The maximum rate at which the oscillating PNS can radiate energy in acoustic 
waves is 
$\dot{E} \sim E_{\mathrm{sat}} T^{2}/t_{\mathrm{cross}} \sim 10^{51}\,\ergpers$ 
for a transmission coefficient $T_{\ell}^{2} = 0.5$. As a consequence, our 
computed energy fluxes are at the limit where non-linear coupling may affect 
our results, but we do not expect non-linear suppression far below our 
estimates.

It is worth noting that the wave crossing time is only a few times larger than  
the wave oscillation periods, and comparable to the daughter mode periods.      
Hence, we consider it unlikely that non-linear coupling can dissipate the waves 
faster than the wave crossing time, and thus unlikely it can strongly attenuate 
the gravity waves before they transition into acoustic waves. This is an        
important distinction from lower frequency waves or modes trapped in the inner  
core, which must undergo many more oscillation cycles before their energy      
leaks out into the envelope.

Our estimates of wave heating are lower limits in the sense that they do not    
allow for gravity waves to reflect multiple times within the PNS before         
tunnelling into outgoing acoustic waves. In the absence of damping, multiple     
reflections could allow the acoustic wave energy flux to approach $\lwave$ 
(Equation~\ref{lwaves}) rather than $\lwave\,T_{\ell}^{2}$. Based on the 
calculations above, however, non-linear coupling likely can prevent gravity 
waves from reflecting many times and accumulating energy in the outer PNS. With 
this in mind, it is unlikely that multiple reflections would greatly increase 
the wave heating rates.

\subsection{Impact of waves on shock revival}
\label{subsec:sasi}
\begin{figure}                                                                  
\centering                                                                      
\includegraphics[width=0.97\columnwidth]{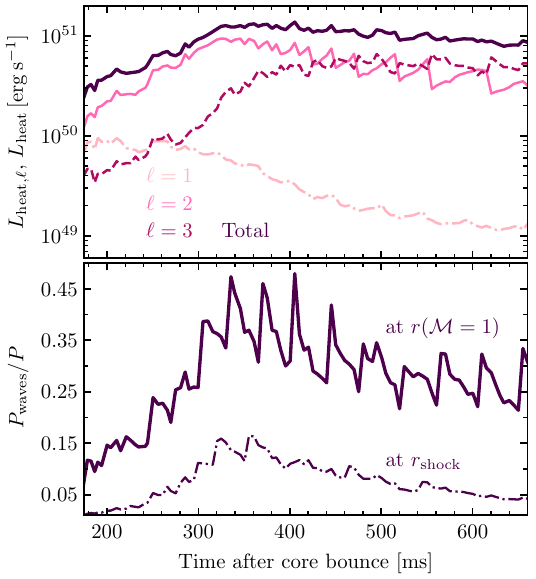}
\caption{(Top panel) Temporal evolution of the net heating rates corrected for  
neutrino damping $\lheatl$ in $\ell=1$ (dash-dotted light pink line), $\ell=2$ 
(solid hot pink line), and $\ell=3$ (dashed deep pink line) waves. The total net 
heating rate $\lheat$, summed over excited modes, is overlaid with a deep 
purple line. (Bottom panel) The ratio of the wave pressure (summed over excited 
modes) to the total pressure $P_{\mathrm{waves}}/P$ as a function of time, as 
measured where the Mach number $\mathcal{M}=1$ (thick deep purple line), and 
immediately below the shock (thin dot-dashed deep purple line).}                                       
\label{fig:wave-pressure}                                                       
\end{figure}

Under the assumption that any wave energy lost to neutrinos free-streams out through the shock and escapes, an approximate lower limit on   
the wave energy heating rate in the post-shock region may be obtained simply  
by multiplying the wave transmission rate by the fraction not attenuated by neutrino  
damping. Explicitly, an estimate for the corrected heating rate                 
$\lheat$ is just                                                 
\begin{align}                                                                   
    \lheat &= \sum_{\ell=1}^{\ellc}\ltrans\,\exp\left[-x_{\ell}\right]\,,\notag\\
    &=  \sum_{\ell=1}^{\ellc} \lheatl\,.                   
\end{align}
In the top panel of Fig.~\ref{fig:wave-pressure}, we show the wave energy       
heating rate in the post-shock region as a   
function of time for $\ell = 1-3$ waves, in addition to the total heating
rate summed over all excited angular modes. While slightly reduced relative to 
Fig.~\ref{fig:heating}, we see that corrected wave heating rates are still 
expected to be substantial, with typical values 
$\lheat \sim 4-10 \times 10^{50}\,\ergpers$ in wave power reaching the shock 
over the course of the simulation.

Exactly what happens as the waves reach the shock is complicated, but in        
general, downward reflection of the waves at the shock is expected. Acoustic    
waves are reflected at the discontinuity in density, entropy, and flow velocity,
due to the supersonic inflow velocity above the shock. Generally speaking,      
acoustic waves here are partially reflected (to inwardly propagating acoustic   
waves) and partially transformed into vorticity and entropy waves through       
perturbation of the shock. Wave amplitudes can be augmented by tapping the      
kinetic energy of mass falling onto the shock, hence we expect the wave         
pressure calculations below to be conservative estimates. The interaction of    
outgoing acoustic waves with the stalled shock has been studied extensively in 
the context of the SASI                             
(see, e.g.,~\citealt{foglizzo:etal:2007,foglizzo:2009,guilet:etal:2010,guilet:foglizzo:2012}),
but the particulars of the wave-shock interaction are beyond the scope of this  
study and left to future work. We also direct the interested reader to work 
by~\citet{abdikamalov:etal:2016}, \citet{huete:etal:2018}, and \citet{abdikamalov:etal:2018} for       
discussions on the interaction of waves with the stalled shock.

The momentum flux carried by the acoustic waves is imparted onto the shock as       
they reflect at it, just as photons impart momentum when they reflect at a surface. 
The waves therefore exert pressure on the shock, and 
a useful figure of merit to consider is the ratio of wave pressure              
$P_{\mathrm{waves}}$ to total fluid pressure at the shock. The momentum flux in 
acoustic waves is $\lwave/(2 c_{s})$, and thus the wave pressure can be    
estimated from the heating rate;                                                
\begin{equation}                                                                
    P_{\mathrm{waves}} = \frac{\lheat}{4\pi r^{2}c_{s}}\,,
\end{equation}                                                                  
where we employ the energy transport rate as computed above, corrected for the  
effects of neutrino damping.

In the lower panel of Fig.~\ref{fig:wave-pressure}, we show the ratio of wave 
pressure to total pressure $P_{\mathrm{waves}}/P$ as a function of time. We 
show this quantity as measured in two places; just exterior to the  
entropy discontinuity marking the shock (a lower bound), in addition to the     
wave turning point where $\mathcal{M} = 1$ (an upper bound). Below the entropy  
discontinuity, we see wave pressure may account for over $15$ per cent of the  
total pressure after $325\,\mathrm{ms}$ post-bounce, falling to $\sim5$ per cent 
by the end of our simulation. In contrast, where $\mathcal{M} = 1$, in  
excess of $20$ per cent of the total pressure is seen in waves from early times,       
rising to a maximum of $\sim30-45$ per cent of the total pressure contributed in waves   
between $300$ and $450\,\mathrm{ms}$ post-bounce. Between $500\,\mathrm{ms}$ through  
the end of the simulation, wave pressure ratios around $\sim25$ and $35$ per cent are          
consistently seen. We believe these estimates are relatively conservative, as 
wave pressure behind the shock may build up over time as a consequence of 
multiple interactions facilitated by wave reflection and/or multiple 
advective-acoustic cycles, similar to the SASI.

\section{Discussion}
\label{sec:discussion}

\subsection{Implications for CCSN explosions}
It is important to distinguish our work from the so-called acoustic mechanism   
\citep{burrows:etal:2006:a,burrows:etal:2006:b}, as also discussed in           
Section~\ref{sec:basicidea}. In the acoustic mechanism, PNS oscillations are       
driven by asymmetric accretion onto the outer PNS, or by the SASI               
\citep{yoshida:etal:2007}. These mechanisms essentially transfer kinetic energy 
from the accretion flow into the PNS and then back out toward the gain region,  
so there is no net transfer of energy from the PNS core outward. In our         
mechanism, wave energy excited by core convection is nearly independent of the  
accretion rate or asymmetry. Unlike the acoustic mechanism, the mechanism       
considered here can draw from the larger reservoir of PNS core binding energy   
($E\sim 5 \times 10^{53}\,\mathrm{erg}$), albeit inefficiently. More            
importantly, wave excitation can persist for as long as the inner convection    
zone exists, which is likely for several seconds after core                     
bounce~\citep{burrows:1987,roberts:2012}. Consequently, wave power generated in 
this scenario could contribute to the explosion power for long periods of       
time, even after the explosion has been fully launched and accretion power is   
negligible.

The extent to which acoustic energy aids explosion energy was studied in detail 
by \cite{harada:etal:2017}. On its own, acoustic power nearing 
$10^{52}\,\mathrm{erg}$ is required to drive an explosion, which is unlikely to be 
generated by PNS core convection. However, in realistic models with neutrino 
luminosities of several $10^{52}\,\ergpers$, less than $10^{51}\,\ergpers$ of 
wave power could have a large effect, especially at late times when the mass 
accretion rate has declined. Hence, our predicted wave energy fluxes of 
$\sim10^{51}\,\ergpers$ could play a pivotal role in shock revival for some 
supernovae.

Another important effect of waves may be late time (beyond $\sim 500$ ms)       
energy deposition in the PNS wind. Once the shock has been driven to large      
radii, the waves will steepen into shocks as they propagate into the            
low-density PNS wind (see \citealt{roberts:2012:thesis}), thereby thermalizing 
their energy in the inner explosion ejecta. Extrapolating 
Fig.~\ref{fig:wave-pressure} to late times, wave power of 
nearly $10^{51}\,\ergpers$ may be sustained for more than a second after bounce, 
potentially contributing as much as $\sim \! 10^{51}\,\mathrm{erg}$ to the 
explosion energy. While some of this energy may well be lost to neutrino 
cooling, late-time wave energy deposition could significantly contribute to the 
final explosion energy. In light of the very low explosion energies 
($E_{\mathrm{exp}}\sim\mathrm{few}\times 10^{50}\,\mathrm{erg}$) currently      
reached at end of many CCSN simulations, this effect seems especially           
important. The extent to which wave effects will contribute to the explosion    
energy at late times will depend on how efficiently convectively excited        
gravity waves are transmitted into outgoing acoustic waves. This efficiency     
is dependent on the structure of the outer PNS at late times, which will be     
quite different in the case of a successful explosion. As our simulations, which
do not explode, cannot be used to compute this efficiency, we hope to quantify  
late time wave heating rates in future work.

A crucial implication of our study is that CCSN simulations that do not         
resolve the dynamics within the inner $20\mathkm$ of the PNS may be missing an 
important source of explosion energy because they will not capture the effects 
of waves excited by the inner PNS convection zone. Indeed, many 3D and/or 
long-running simulations 
(see, e.g.,~\citealt{mueller:2015,bmueller:2017,mueller:etal:2019}) evolve the 
inner PNS in spherical symmetry to allow for longer time-steps to be taken. 
The impact of this cut on the ability of a given simulation 
to resolve PNS convection and associated wave heating is dependent on the 
boundary criterion employed. Often, a density cut at 
$\rho\sim5\times10^{11}\,\mathrm{g}\,\mathrm{cm}^{-3}$ is chosen, such that 
higher densities are evolved in spherical symmetry. As can be seen from 
Fig.~\ref{fig:kippenhahn}, this corresponds to a radial cut exterior to 
$\sim20\,\mathkm$ and, as such, simulations employing this technique would 
miss the effect of waves excited by PNS convection and associated heating. Other 
simulations employing instead an explicit radius cut (often around 
$\sim10\,\mathkm$) should indeed resolve some of the convection and waves, but 
it is not clear how the dynamics would be affected by imposing an artificial 
inner boundary to the PNS convective zone, particularly at later times.  
Other simulations (e.g.,~\citealt{burrows:etal:2019:a,vartanyan:etal:2019}) do    
resolve the dynamics of the inner PNS, though it is still not clear whether the 
$0.5\mathkm$ resolution is sufficiently fine to accurately capture the    
effects of convective wave excitation and propagation. To help quantify the     
effects of waves, it would be interesting to compare simulations that evolve the
inner PNS both with and without spherical symmetry. If waves are contributing   
to the explosion, we expect simulations that do resolve the dynamics of the     
inner PNS to explode more easily and with larger energies.

Energy fluxes from non-radial oscillations can be computed in simulations via   
\begin{equation}                                                                
    \lwave = \int d S \, \delta P \, \delta v_{r}\,.                 
\end{equation}                                                                  
Here, the integral is taken over a spherical surface area $S$, while $\delta P$ 
and $\delta v_{r}$ are the non-radial components of the pressure and radial       
velocity perturbations, respectively. Care should be taken to distinguish the   
upward wave energy flux from the downward energy and mass flux due to 
accretion. \cite{mueller:2015} examined the \textit{downward} energy flux of 
gravity waves that are excited by the \textit{outer} convection zone, finding in 
2D simulations downward energy fluxes as large as a few 
$\times 10^{50}\ergpers$, though their 3D simulations find much smaller downward 
energy fluxes due to less vigorous and less coherent convective motions. 
However, as those simulations did not resolve the inner PNS, the upward energy 
flux due to gravity waves excited by the inner PNS convection zone would have 
been missed. In general, both types of waves are expected be present.

Gravity waves excited by PNS convection may be present in some simulations even 
if it is not recognized or discussed in corresponding papers. We expect         
non-radial horizontal motions of $\sim \! 10^{3}-10^{4}\kmpers$ (with somewhat 
smaller radial motions) due to gravity waves in the outer stably stratified 
region of the PNS, with larger motions at larger radii where the material is 
less dense. In the outer gravity wave cavity, the waves of interest have only 
one or two nodes and, hence, they may not appear particularly wavelike in 
simulations, perhaps instead resembling a large-scale ringing of the PNS. We 
note that gravity waves have opposite radial group and phase velocities, hence 
waves that carry energy upward have phases that propagate \textit{downward}, 
which could easily be mistaken for downward propagating waves excited by 
convection and non-radial accretion from the overlying gain region. It is also 
possible that some of the vigorous motion in the gain region, usually attributed to 
neutrino-driven convection, could be caused partially by acoustic waves 
emerging from below.

Using simulations that resolved PNS convection, \cite{andresen:etal:2017} found 
that inner PNS convection does indeed excite waves (although they did not       
quantify the hydrodynamic energy flux), and that these PNS motions can          
contribute to the gravitational wave (GW) emission from CCSNe. Using the same   
runs as \cite{burrows:etal:2019:a}, \cite{radice:etal:2018} found that motions in 
the gain region helped excite fundamental and g-mode oscillations of the PNS, 
which dominated the GW emission found by their simulations. 
\cite{torres-forne:etal:2019:a,torres-forne:etal:2019:b} found very similar 
results. The peak GW frequencies they find are similar but slightly larger than 
the gravity waves we expect to be excited by convection in the PNS mantle. As 
fundamental PNS modes are excited by waves from the outer convection zone 
(behind the shock) at early times (from $\sim \! 100\,\mathrm{ms}$ post-bounce), 
this higher frequency emission dominates the GW spectrum because the energy flux 
in GWs is strongly dependent on frequency 
($\dot{E}_{\mathrm{GW}} \propto \omega^6$). At late times, after neutrino 
heating has died down, it is possible that PNS oscillations excited by PNS 
convection (which will persist for several seconds as the PNS cools and 
contracts) could dominate the GW emission. Indeed, if true, this could mirror 
the scenario we are proposing in this paper; at early times, neutrino heating 
may dominate the explosion dynamics (and the GW signature), but waves excited 
by convection in the PNS mantle could contribute significantly to explosion 
dynamics, PNS evolution, and consequently the GW emission, beyond a few hundred 
milliseconds post-bounce. Further investigations into wave heating from PNS 
convection will require simulations that resolve the PNS to evolve it from 
formation out to several seconds after core bounce.

Ours is not the first work to consider the impacts of convectively excited waves
on CCSNe. \cite{metzger:etal:2007} examined the impacts of convectively excited 
Alfv{\'e}n waves on $r$-process nucleosynthesis in PNS winds. Wave heat         
deposited at the base of the PNS wind can help drive it outward more rapidly,   
such that it maintains lower $Y_{e}$ and produces more $r$-process elements.    
\cite{metzger:etal:2007} found Alfv{\'e}n wave heating may be important, but    
only for protomagnetar-type remnants with magnetic fields on the order 
$B \sim 10^{15}\,\mathrm{G}$. Even then, they estimated wave heating rates of  
only $L \lesssim 10^{49}\ergpers$, roughly two orders of magnitude smaller 
than the wave heating estimates we predict here. \cite{suzuki:etal:2008} found 
that Alfv{\'e}n waves could revive the shock only for remnants with exceptionally 
large field strengths ($B \gtrsim2\times 10^{15}\,\mathrm{G}$) and perturbation 
amplitudes. \cite{roberts:2012:thesis} considered the impact of gravito-acoustic
waves excited by PNS convection on nucleosynthesis in neutrino driven winds. 
There, it was found that energy deposition by gravito-acoustic waves could 
strongly impact the dynamics of the wind, but the impact on nucleosynthesis 
depended strongly on the radius at which the acoustic waves steepened into 
shocks. Under favourable conditions, this extra heating resulted in production 
of the $r$-process in mildly neutron-rich conditions. Based on that work and 
the results found here, we believe that waves hydrodynamically excited by 
PNS convection are usually more important for both the explosion and 
nucleosynthesis in the neutrino driven wind than Alfv{\'e}n waves. In light of 
our results, the effect of wave heating on nucleosynthesis in the 
neutrino-driven wind should be reinvestigated.

\subsection{Uncertainties}
As explained in Section~\ref{sec:wavegen}, we make the approximation that all wave 
flux excited by PNS convection is at the convective overturn frequency~$\wcon$  
and that the angular wave spectrum is flat in modes $\ell\in[1,\ldots,\ellc]$.  
While the spectral behaviour of waves excited in this way is not well 
understood, our prescription is rooted in the assertions made       
in~\citet{goldreich:kumar:1990}, \citet{kumar:etal:1999}, and \citet{lecoanet:quataert:2013}, on the  
basis of bulk Reynolds stresses driving convective wave excitation. The         
propagation path of the excited waves, as shown in Fig.~\ref{fig:propagation},  
is clearly frequency dependent, a statement we have quantified with the         
fraction of wave flux transmitted from gravity waves to acoustic waves in the   
outer PNS as a function of frequency shown in Fig.~\ref{fig:transmission}.      
Based on these figures, it would appear that emission at higher frequencies, as 
argued for by~\citet{rogers:etal:2013} and \citet{pincon:etal:2016} on the basis of plume   
incursion driving convective wave excitation, may slightly increase the         
fraction of transmitted wave flux. As a consequence, despite their somewhat     
crude and uncertain nature, we believe our estimates of the wave flux excited   
by PNS convection are conservative. More importantly, we see from 
Fig.~\ref{fig:propagation} that the transmitted wave flux may be higher or 
lower at different wave frequencies for different values of $\ell$, but 
integrating over a broad frequency spectrum will yield wave fluxes of similar 
magnitude to our rough estimates.

Neutrino damping likely has a substantial impact on wave heating rates. As we   
have shown in Fig.~\ref{fig:nudamp}, a moderate fraction of the wave flux is    
likely to be damped away through increased neutrino energy losses in the        
marginally optically thin region around the neutrinosphere. Our  
estimates of neutrino damping, admittedly somewhat crude, use a WKB 
approximation for the waves (which is only marginally valid for low-order       
gravity waves), and do not accurately predict neutrino losses where the         
optical depth across a wavelength is near unity. Hence, a more careful          
assessment of neutrino damping of waves from CCSN simulations is warranted.

Given all emission at $\wcon$ with a flat angular spectrum in                   
$\ell\in[1,\ldots,\ellc]$, strong non-linear effects are likely to be important 
only for gravity waves in the PNS core at early times, and perhaps additionally 
for acoustic waves in the gain region on very close approach to the shock.      
While non-linear damping in the core may prevent wave energy emitted into the   
core from escaping, the estimates for $\ltrans$ we present consider only 
outwardly propagating waves, and thus can only be increased by wave flux 
escaping from the inner core. Our estimates suggest that non-linear three-mode  
coupling, as discussed by \cite{weinberg:quataert:2008}, is marginally          
important for the waves we consider. Waves with lower $\omega$ and higher-order 
angular wavenumber $\ell$ are more susceptible to non-linear effects due to     
their larger radial wavenumbers $|k_{r,\ell}|$ and slower group velocities      
$v_{\mathrm{gr}}$. If convection excites waves at predominantly lower           
frequencies (or higher $\ell$ than our estimates), three-mode coupling could    
reduce the wave flux transmitted into acoustic waves. Alternatively, convective 
excitation at higher frequency or lower $\ell$ could increase the transmitted   
wave flux. In the latter case, increased transmission of acoustic wave flux     
will increase the non-linearity of acoustic waves in the gain region, increasing
the likelihood of wave damping through weak shock formation.

Beyond the points already highlighted here and previously alluded to, a crucial 
limitation underscoring this work is the failure of our simulations to          
explode. The thermodynamic structure of the accreting PNS, particularly in the  
immediate post-shock region, will be altered in the case of a successful        
explosion. Although this is unlikely to impact the development or persistence   
of convection in the PNS mantle over the time-scales considered in this study,   
the propagation of waves through the outer PNS and the gain regions is likely   
to be impacted. A quantitative evaluation of the effect on wave pressure and    
corrected heating rates are beyond the scope of this study, but should be       
considered in future work.

\section{Conclusion}
\label{sec:conclusion}
In the first few seconds after core collapse, energy transport by convectively  
excited waves from the inner PNS toward the shock may have a substantial impact 
on the outcome of core collapse and explosion energy. To quantify wave energy   
transport, we have modelled the core collapse of a $15\,\msun$ progenitor and   
followed post-bounce evolution for $660\,\mathrm{ms}$. We used a                
spherically symmetric, general relativistic hydrodynamics code employing a      
mixing length theory prescription for 
convection~\citep{roberts:2012:thesis,roberts:etal:2012} to estimate wave          
generation rates, analyse wave propagation within the PNS, and compute wave     
energy fluxes behind the stalled shock.

Convection develops in the PNS mantle after around $200\,\mathrm{ms}$ 
post-bounce due to deleptonization and entropy changes as the PNS contracts. We 
see convective luminosities of $\lcon \! \sim 10^{53}\ergpers$ across the 
inner convective region, of which a few $10^{51}\ergpers$ is   
expected to be radiated outwards as gravity waves from the outer edge with      
frequencies near the convective turnover frequency, which we estimate to        
increase from $\sim10^{3}$ to $\sim4\times10^{3}\radpers$ over the 
course of the simulation. Because the convectively excited gravity waves 
encounter a relatively narrow evanescent region between the PNS and the gain 
region, a large fraction ($T_{\ell}^{2} \sim 1/3$) of their energy is 
transmitted into acoustic waves that propagate out toward the shock. In the 
post-shock region, we find net wave energy transport rates slightly exceeding 
$10^{51}\ergpers$ sustained through the end of the simulation.

Neutrino damping of acoustic waves in the outer PNS is likely to be 
significant, with a moderate fraction of wave energy dissipated due to 
escaping neutrinos. Accounting for the effects of neutrino damping, we still 
find wave heating rates of nearly $10^{51}\ergpers$ sustained through the end 
of the simulation (c.f.~Fig.~\ref{fig:wave-pressure}). While we do not expect 
non-linear effects to drastically alter our results, steepening of acoustic 
waves could cause energy deposition in the post-shock region, especially once 
the explosion commences and the shock moves outward. Non-linear three-mode 
coupling in the outer PNS could moderately reduce the amount of wave energy 
escaping toward the gain region, particularly if the spectrum of the waves 
excited by convection peaks at lower frequencies than we have assumed, though we 
find it unlikely to drastically affect our results.

Although we do not expect wave energy transport to be the primary driver of the  
supernova explosion, our study here shows that its impact is expected to be     
highly significant, contributing as much as $40$ per cent of the pressure upon the     
shock. Since many configurations of core collapse exist very close to the       
threshold between collapse and explosion                            
(see, e.g.,~\citealt{melson:etal:2015:b, bmueller:etal:2017}), waves may play a   
crucial role in numerous events. Furthermore, since wave heating    
extends to late times (beyond $500\,\mathrm{ms}$ post-bounce) and falls off     
at a slower rate than neutrino heating, wave energy may significantly 
contribute to the final energy of the explosion. Future simulations can better  
quantify the impact of waves, and we encourage CCSN modellers to attempt to 
resolve the inner PNS in order to capture wave excitation and propagation 
originating from convection in the PNS mantle. We also suggest a re-examination 
of the outgoing hydrodynamic energy flux from the PNS, as convectively excited 
waves could be present in some existing simulations, but may be unrecognised 
or even mistaken for downgoing waves.

The physics governing CCSNe is rich and diverse, encompassing turbulent 
hydrodynamic instabilities and complex radiative transfer of neutrinos. In the 
absence of unlimited computational resources, the extent to which these systems 
can be modelled is constrained by the approximations used to make such studies 
viable. Wave physics in itself is a complex field which is not fully 
understood, and thus our results are both estimative and subject to a number of 
uncertainties that have been outlined in this study and can be improved upon 
with further investigation. Nevertheless, it is clear that the accurate 
quantification of the impact of wave heating from PNS convection on the revival 
of the stalled supernova shock is vital to developing a comprehensive 
understanding of the CCSN explosion mechanism.

\section*{Acknowledgements}
We acknowledge helpful discussions with T. Foglizzo, A. Harada, and E. Lentz of 
great benefit to this work. SEG thanks S. Nissanke, R. Wijers, the GRavitation
AstroParticle Physics Amsterdam (GRAPPA) group, and the Anton Pannekoek 
Institute (API) for their hospitality at Universiteit van Amsterdam (UvA), where 
much of this work was carried out. This research is funded in part by an 
Innovator Grant from The Rose Hills Foundation, the Sloan Foundation through 
grant FG-2018-10515, and by the National Science Foundation under 
Grant No. NSF PHY-1748958. 

\textit{Software}: All figures presented in this paper were produced using 
\textsc{Python}, \textsc{Matplotlib}~\citep{matplotlib}, 
\textsc{NumPy}~\citep{numpy1,numpy2}, and \textsc{SciPy}~\citep{scipy}.




\bibliographystyle{mnras}
\bibliography{refs} 



\appendix

\section{Wave damping via neutrino dissipation}
Wave energy dissipation is expected to occur due to non-adiabatic corrections 
from neutrino emission. The energy loss rate per unit mass due to neutrino 
damping, $\dot{\varepsilon}_{\nu}$, can be calculated as
\begin{align}
    \dot{\varepsilon}_{\nu} &= \delta T\, \delta \left( \frac{\mathrm{d}s}{\mathrm{d}t} \right)_{\nu}\,,
\end{align}
where $\delta$ denotes the Lagrangian variation of the following term, and 
\begin{align}
\label{dsdt}
    \left( \frac{ds}{dt} \right)_{\nu} &= \sum_{\nu_{i} \in \{\nu_{e}, \bar{\nu}_{e}, \nu_{x}\}} \left( \frac{S_{\nu_{i}}}{T} - \frac{1}{n_{B}T} \left[ \vec{\nabla} \cdot \vec{H}_{\nu_{i}} - \mu_{\nu_{i}} \vec{\nabla} \cdot \vec{F}_{\nu_{i}}  \right] \right)  \,
\end{align}
is the rate of change of the specific entropy $s$ due to neutrino losses. 
Here, the source function $S_{\nu_{i}}$, number flux $F_{\nu_{i}}$, energy 
flux $H_{\nu_{i}}$, and chemical potential $\mu_{\nu_{i}}$ are distinct for 
each neutrino species $\nu_{i}$. In equations (\ref{dsdt}) and (\ref{snu}), 
Boltzmann's constant $k_B \rightarrow 1$. The source function, which encompasses effects 
from both the emission and absorption of neutrinos of species $\nu_{i}$, is 
defined
\begin{align}
\label{snu}
    S_{\nu_{i}} = &\frac{4\pi c}{n_{B}} \int_{0}^{\infty} 
    \frac{d\omega}{ (2\pi \hbar c)^{3}}\,\omega^{3}\,
    \kappa^{*}_{\nu_{i}}(\omega) \notag\\
    &\times\left[ f_{\nu_{i},\mathrm{eq}} (\omega, T, \mu_{\nu_{i},\mathrm{eq}}) - f_{\nu_{i}}(\omega) \right] \,,
\end{align}
where $\kappa^{*}_{\nu_{i}}$ is the stimulated absorption opacity (with units
of inverse length), $f_{\nu_{i}}$ is the neutrino distribution function, and 
$f_{\nu_{i},\mathrm{eq}}$ is the distribution function describing a 
hypothetical population of $\nu_{i}$ neutrinos in thermal and chemical 
equilibrium with the background medium, which takes the usual Fermi-Dirac 
form. 

Generally speaking, the distribution functions for each neutrino population 
must be calculated from the Boltzmann equation. In the optically thin and 
optically thick limits, however, one can make approximations and compute 
$\dot{\varepsilon}_{\nu}$ in terms of local thermodynamic quantities 
and their spatial gradients. We now discuss these cases in turn.

\subsection{Optically thick limit}
\label{app:thick}
In the optically thick (i.e. diffusive) limit, 
$f_{\nu_{i}} \rightarrow f_{\nu_{i},\mathrm{eq}}$ plus some small 
anisotropic correction. As such, neutrino emission and absorption are 
locally balanced and $S_{\nu_{i}} \rightarrow 0$.

As derived in~\cite{pons:etal:1999}, the diffusive fluxes in this limit for 
neutrino species $\nu_{i}$ may be written in terms of the local temperature and 
neutrino degeneracy parameter 
($\eta_{\nu_{i}} = \mu_{\nu_{i}}/T$) gradients:
\begin{align}
    \vec{F}_{\nu_{i}} &= -\frac{cT^{2}}{6\pi^{2} (\hbar c)^{3}} \left[ D_{3,\,\nu_{i}} \vec{\nabla}T  + D_{2,\,\nu_{i}} \,T\,\vec{\nabla}\eta_{\nu_{i}}\right]\,,\\
    \vec{H}_{\nu_{i}} &= -\frac{cT^{3}}{6\pi^{2} (\hbar c)^{3}} \left[ D_{4,\,\nu_{i}} \vec{\nabla}T  + D_{3,\,\nu_{i}} \,T\,\vec{\nabla}\eta_{\nu_{i}} \right]\,,
\end{align}
where diffusion coefficients $D_{n,\nu_{i}}$ (essentially inverse Rosseland 
mean opacities) are defined
\begin{align}
    D_{n,\nu_{i}} &= \int_{0}^{\infty} d\omega\,\frac{ \omega^{n} f_{\nu_{i},\mathrm{eq}} (\omega)\left[ 1 - f_{\nu_{i},\mathrm{eq}} (\omega) \right] }{ T^{n+1} \left[ \kappa^{*}_{\mathrm{abs},\nu_{i}}(\omega) + \kappa^{*}_{\mathrm{scat},\nu_{i}}(\omega) \right]}\,,
\end{align}
and $\kappa^{*}_{\mathrm{scat},\nu_{i}}(\omega)$ here is the
scattering transport opacity for neutrino species $\nu_{i}$, with units of 
inverse length.

Diffusive losses here are dominated by neutral-current scattering of 
heavy-lepton neutrinos on nucleons. As the heavy-lepton chemical potential 
$\mu_{\nu_{x}} = 0$ everywhere, we see that
\begin{align}
    \left( \frac{\mathrm{d}s}{\mathrm{d}t} \right)_{\nu} &\approx -\frac{1}{n_{B}T}\vec{\nabla} \cdot \vec{H}_{\nu_{x}}\,.
\end{align}
Using the diffusive approximations introduced above, the heavy-lepton 
neutrino energy flux $\vec{H}_{\nu_{i}}$ may be written
\begin{align}
    \vec{H}_{\nu_{x}} &= -\frac{4 c T^{3}}{6\pi^{2} (\hbar c)^{3}}\,D_{4,\,\nu_{x}} \vec{\nabla} T\,,
\end{align}
where the factor of four here in the numerator accounts for the summed flux 
over the four flavours encompassed by the heavy-lepton neutrino species, 
$\nu_{x} \in \{ \nu_{\mu},\,\bar{\nu}_{\mu},\,\nu_{\tau},\,\bar{\nu}_{\tau} \}$. 

To compute $D_{4,\,\nu_{x}}$, we consider the transport opacity for heavy-lepton 
neutrinos from neutral-current scattering on nucleons. Employing 
expressions from Sections 3.4 and 3.5 of~\citet{burrows:etal:2006:a}, we may write
\begin{align}
\kappa^{*}_{\mathrm{scat},\nu_{x}} ( \omega) \approx n_{B} \sigma_{0} \left( \frac{\hbar \omega}{m_{e}c^{2}} \right)^{2} C(Y_{e})\,.
\end{align}
Here, $\hbar \omega$ is the neutrino energy and $C(Y_{e})$ encompasses coupling constants and composition effects:
\begin{align}
    C(Y_{e}) &=  \frac{1 + 5g_{A}^{2}}{6} \frac{1-Y_{e}}{4} + 
    \frac{(C^{'}_{V} - 1)^{2} + 5g_{A}^{2} (C^{'}_{A} - 1)^{2} }{6} Y_{e}\,.
\end{align}

Substituting this into the expression for $D_{4,\,\nu_{x}}$ and making the 
change of variables $\hbar \omega \rightarrow x = \omega/T$ yields
\begin{align}
    D_{4,\,\nu_{x}} &\approx \frac{ \pi^{2}(m_{e}c^{2})^{2} }{6\, n_{B}\, \sigma_{0} C(Y_{e}) T^{2}}\,,
\end{align}
where we employ the identity
\begin{align}
    \int_{0}^{\infty}dx\,\frac{x^{2} e^{x}}{(e^{x} + 1)^{2}} = \frac{\pi^{2}}{6}\,.
\end{align}

The energy flux in heavy-lepton neutrinos naturally follows
\begin{align}
    \vec{H}_{\nu_{x}} &\approx -\frac{c \,(m_{e}c^{2})^{2}}{9\, (\hbar c)^{3} \sigma_{0}} 
    \frac{ T }{ n_{B} C(Y_{e}) } \vec{\nabla}T \,,
\end{align}
from which we find
\begin{align}
    \left( \frac{\mathrm{d}s}{\mathrm{d}t} \right)_{\nu} &\approx \frac{1}{n_{B}T}\,
     \frac{c\, (m_{e}c^{2})^{2}}{9\, (\hbar c)^{3}\, \sigma_{0}} \notag\\
    &\times \left[ \vec{\nabla} \left(  \frac{T}{n_{B}C(Y_{e})} \right) \cdot \vec{\nabla}T + \frac{T}{n_{B}C(Y_{e})} \nabla^{2} T  \right]\,.
\end{align}

Taking now the Lagrangian variation of this expression and neglecting terms 
second-order and above in perturbations, we can write
\begin{align}
    \delta \left( \frac{\mathrm{d}s}{\mathrm{d}t} \right)_{\nu} &\approx
    \frac{c\, (m_{e}c^{2})^{2}}{9(\hbar c)^{3} \sigma_{0} } 
    \frac{\nabla^{2}\delta T}{n_{B}^{2} C(Y_{e})}\,,\notag\\
    &\approx -\frac{c\, (m_{e}c^{2})^{2}}{ 9\, (\hbar c)^{3}\, \sigma_{0} } 
    \frac{k^{2}\delta T}{n_{B}^{2} C(Y_{e})}\,,
\end{align}
where we have employed the WKB approximation to substitute 
$\nabla^{2} \delta T \rightarrow -k^{2} \delta T$ and to ignore terms of lower order in $k$.

From here, the damping rate in neutrinos in the optically thick limit is just
\begin{align}
    \dot{\varepsilon}_{\nu,\,\mathrm{thick}} &= \delta T \, \delta\left( \frac{ds}{dt} \right)_{\nu}\,,\notag\\
    &\approx -\frac{c\, (m_{e}c^{2})^{2}}{9\,(\hbar c)^{3}\,\sigma_{0}} 
    \frac{k^{2}\,\delta T^{2}}{n_{B}^{2}\,C(Y_{e})}\,.
\end{align}

\subsection{Optically thin limit}
\label{app:thin}
In the optically thin limit, $f_{\nu_{i}}(\omega)\approx 0$, and the 
entropy evolution equation takes the simplified form
\begin{align}
    \left(\frac{\mathrm{d}s}{\mathrm{d}t}\right)_{\nu} &=
    \sum_{\nu_{i}\in\{\nu_{e},\,\bar{\nu}_{e},\,\nu_{x}\}}
    \frac{\dot{\epsilon}_{\nu_{i}}}{T}\,,
\end{align}
where the neutrino emission rates $\dot{\epsilon}_{\nu_{i}}$ 
(with units of per second per baryon) are defined
\begin{align}
    \dot{\epsilon}_{\nu_{i}} &= \frac{\hbar}{2 \pi^2 n_{B} c^2}
    \int_{0}^{\infty} d\omega 
    \frac{\omega^{3}\,\kappa^{*}_{\nu_{i}}(\omega)}{e^{\hbar \omega/T - \eta_{\nu_{i}}} + 1}\,.
\end{align}

In this limit, losses are dominated by emission of $\nu_{e}$ and 
$\bar{\nu}_{e}$ neutrinos, driven by charged-current capture on nucleons.  
Explicitly, the relevant stimulated absorption opacities may be written
\begin{align}
    \kappa^{*}_{\nu_{e}}(\hbar \omega) &= n_{B} (1- Y_{e})\,\sigma_{0}
    \left(\frac{1 + 3g_{A}^{2}}{4}\right)\,
    \left(\frac{\hbar \omega + \Delta_{np}}{m_{e}c^{2}}\right)^{2} \notag\\
    &\times \left[ 1 + e^{-(\hbar \omega/T - \eta_{\nu_{e},\mathrm{eq}})}  \right]\,,\\
    \kappa^{*}_{\bar{\nu}_{e}}(\hbar \omega) &= n_{B}\,Y_{e}\,\sigma_{0}
    \left(\frac{1 + 3g_{A}^{2}}{4}\right)\,
    \left(\frac{\hbar \omega - \Delta_{np}}{m_{e}c^{2}}\right)^{2} \notag\\
    &\times \left[ 1 + e^{-(\hbar \omega/T + \eta_{\nu_{e},\mathrm{eq}})}  \right]\,,
\end{align}
where $\Delta_{np} = (m_{n} - m_{p})\,c^{2}$ is the energy-equivalent mass 
difference between neutrons and protons. Substituting these expressions 
into the formulae above and, once more, making the variable substitution 
$\omega \rightarrow x = \hbar \omega/T$, we find a total emissivity of the form
\begin{align}
    \dot{\epsilon}_{\nu_{e}+\bar{\nu}_{e}} &\approx 
    \frac{c\sigma_{0}}{8\pi^{2} (\hbar c)^{3} (m_{e}c^{2})^{2} }\,
    C_{\nu_{e}+\bar{\nu}_{e},\,\mathrm{thin}}\,T^{6}\,,
\end{align}
where we have defined $C_{\nu_{e}+\bar{\nu}_{e},\,\mathrm{thin}}$, which encompasses the 
relevant coupling constants and compositional effects, as
\begin{align}
    C_{\nu_{e}+\bar{\nu}_{e},\,\mathrm{thin}} &= (1+3g_{A}^{2}) \notag\\ 
    &\times \bigg[ Y_{e} \int_{\Delta_{np}/T}^{\infty} dx\,x^{5} \left(1 - \frac{\Delta_{np}}{xT}\right)^{2} \frac{1 + e^{-x - \eta_{\nu_{e},\mathrm{eq}}}}{1+ e^{x + \eta_{\nu_{e},\mathrm{eq}}}} \notag\\
    +  &(1- Y_{e}) \int_{0}^{\infty} dx\,x^{5} \left(1 + \frac{\Delta_{np}}{xT}\right)^{2} \frac{1 + e^{-x + \eta_{\nu_{e},\mathrm{eq}}}}{1+ e^{x - \eta_{\nu_{e},\mathrm{eq}}}} \bigg]\,,
\end{align}

Returning now to calculate the energy loss rate, we have
\begin{align}
    \dot{\varepsilon}_{\nu_{e}+\bar{\nu}_{e},\,\mathrm{thin}} &\approx 
    \delta \left(\frac{\dot{\epsilon}_{\nu_{e}+\bar{\nu}_{e}}}{T}\right)\,\delta T\,,\notag\\
    &\approx \dot{\epsilon}_{\nu_{e}+\bar{\nu}_{e}}\frac{\delta T}{T}\,
    \bigg[ \left(\frac{\partial \log \dot{\epsilon}_{\nu_{e}+\bar{\nu}_{e}}}{\partial\log T}\right)_{n_{B},Y_{e}} \frac{\delta T}{T} \notag\\
    &\,\,\,\,\,\,\,\,\,\,\,\,\,\,\,\,\,\,\,\,\,\,\,\,\,\,\,\,\,\,+ \left(\frac{\partial \log \dot{\epsilon}_{\nu_{e}+\bar{\nu}_{e}}}{\partial\log n_{B}}\right)_{T,Y_{e}} \frac{\delta n_{B}}{n_{B}} \notag\\
    &\,\,\,\,\,\,\,\,\,\,\,\,\,\,\,\,\,\,\,\,\,\,\,\,\,\,\,\,\,\,+ \left(\frac{\partial \log \dot{\epsilon}_{\nu_{e}+\bar{\nu}_{e}}}{\partial\log Y_{e}}\right)_{T,n_{B}} \frac{\delta Y_{e}}{Y_{e}}\bigg]\,,\notag\\
    &\approx \dot{\epsilon}_{\nu_{e}+\bar{\nu}_{e}}\frac{\delta T}{T}\,\bigg[ 6\,\frac{\delta T}{T} - \frac{\delta Y_{e}}{Y_{e}} \bigg]\,,
\end{align}
Neglecting composition effects to focus instead on the much stronger temperature dependence, we find 
\begin{align}
    \dot{\varepsilon}_{\nu_{e}+\bar{\nu}_{e},\,\mathrm{thin}} &\approx 
    \frac{3c\sigma_{0}}{4\pi^{2}(\hbar c)^{3}(m_{e}c^{2})^{2}}\,C_{\nu_{e}+\bar{\nu}_{e},\,\mathrm{thin}}\,T^{4}\,\delta T^{2}\,.
\end{align}

%
%

\bsp	
\label{lastpage}
\end{document}